\documentclass[12pt,preprint]{aastex}
\usepackage{apjfonts}
\usepackage{natbib}
\usepackage{amssymb, amsmath, amsbsy, epsfig, epsf, slashbox, rotating}

\newcommand{\flux}{erg~s$^{-1}$~cm$^{-2}$}
\newcommand{\lum}{erg~s\ensuremath{^{-1}}}
\newcommand{\lbol}{\ensuremath{L\mathrm{_{bol}}}}

\newcommand{\ledd}{\ensuremath{L\mathrm{_{Edd}}}}
\newcommand{\lratio}{\ensuremath{L/\ledd}}

\newcommand{\lfive}{\ensuremath{\lambda L_{\lambda}(5100}~\AA)}

\newcommand{\msun}{\ensuremath{M_{\odot}}}
\newcommand{\kms}{\ensuremath{\mathrm{km~s^{-1}}}}
\newcommand{\mbh}{\ensuremath{M_\mathrm{BH}}}
\newcommand{\rg}{\ensuremath{r_{\rm \scriptscriptstyle G}}}
\newcommand{\st}{\ensuremath{\sigma_{\rm \scriptscriptstyle T}}}
\newcommand{\massp}{\ensuremath{m_{\rm p}}}

\newcommand{\rs}{\ensuremath{r_{\rm \scriptscriptstyle S}}}
\newcommand{\pnull}{\ensuremath{P_{\mathrm{null}}}}

\newcommand{\chisq}{\ensuremath{\chi^2}}

\newcommand{\ha}{H\ensuremath{\alpha}}
\newcommand{\hb}{H\ensuremath{\beta}}

\newcommand{\hi}{H\,{\footnotesize I}}

\newcommand{\oiii}{[O\,{\footnotesize III}]}

\newcommand{\feii}{{\rm Fe\,{\footnotesize II}}}

\newcommand{\mgii}{Mg\,{\footnotesize II}}

\newcommand{\civ}{C\,{\footnotesize IV}}
\newcommand{\lya}{Ly$\alpha$}

\newcommand{\caii}{Ca\,{\footnotesize II}}

\newcommand{\colnh}{\ensuremath{N_\mathrm{H}}}

\slugcomment{Scheduled to be published in {\it ApJ} in 2011 June}
\shorttitle{\feii\ emission in AGNs}
\shortauthors{Dong et al.}

\begin{document}

\title{What Controls the \feii\ Strength
in Active Galactic Nuclei? }

\author{Xiao-Bo~Dong\altaffilmark{1,2}, Jian-Guo~Wang\altaffilmark{3,4,1,7},
Luis~C.~Ho\altaffilmark{2},
Ting-Gui~Wang\altaffilmark{1}, Xiaohui~Fan\altaffilmark{5,6}, \\
Huiyuan~Wang\altaffilmark{1}, Hongyan~Zhou\altaffilmark{1}, and Weimin~Yuan\altaffilmark{3,4} }

\email{\scriptsize \it Received 2009 February 9; accepted 2011 May 3}

\altaffiltext{1}{Key Laboratory for Research in Galaxies and Cosmology,
The University of Sciences and Technology of China, Chinese Academy of Sciences,
Hefei, Anhui 230026, China; ~ xbdong@ustc.edu.cn}
\altaffiltext{2}{The Observatories of the Carnegie Institution for
Science, 813 Santa Barbara Street, Pasadena, CA 91101, USA; ~ lho@obs.carnegiescience.edu}
\altaffiltext{3}{National Astronomical Observatories, Chinese Academy of Sciences,
Beijing 100012, China; ~ wmy@nao.cas.cn}
\altaffiltext{4}{National Astronomical Observatories/Yunnan Observatory and
Key Laboratory of the Structure and Evolution of Celestial Objects,
Chinese Academy of Sciences, Kunming 650011, China; ~ wangjg@ynao.ac.cn}
\altaffiltext{5}{Steward Observatory, The University of Arizona, Tucson, AZ 85721,
USA; ~ fan@as.arizona.edu}
\altaffiltext{6}{Max Planck Institute for Astronomy, D-69120, Heidelberg, Germany}
\altaffiltext{7}{Graduate School of the Chinese Academy of Sciences,
19A Yuquan Road, P. O. Box 3908, Beijing 100039, China}

\begin{abstract}

We used a large, homogeneous sample of 4178  $z \leq 0.8$ Seyfert 1 galaxies
and QSOs selected from the Sloan Digital Sky Survey to investigate the
strength of \feii\ emission and its correlation with other emission lines and
physical parameters of active galactic nuclei.  We find that the strongest
correlations of almost all the emission-line intensity ratios and equivalent
widths (EWs) are with the Eddington ratio (\lratio), rather than with the
continuum luminosity at 5100~\AA\ ($L_{5100}$) or black hole mass (\mbh);
the only exception is the EW of ultraviolet \feii\ emission, which does
not correlate at all with broad-line width, $L_{5100}$,
\mbh, or \lratio.  By contrast,
the intensity ratios of both the ultraviolet and optical \feii\
emission to \mgii\ $\lambda 2800$ correlate quite strongly with \lratio.
Interestingly, among all the emission lines in the near-UV and optical
studied in this paper (including \mgii\ $\lambda 2800$, \hb, and \oiii\ $\lambda5007$),
the EW of narrow optical \feii\ emission has the strongest correlation with \lratio.
We hypothesize
that the variation of the emission-line strength in active galaxies
is regulated by \lratio\ because it governs
the global distribution of the hydrogen column density
of the clouds gravitationally bound
in the line-emitting region, as well as its overall gas supply. The systematic
dependence on \lratio\ must be
corrected when using the \feii/\mgii\ intensity ratio as a measure of the
Fe/Mg abundance ratio to study the history of chemical evolution in QSO
environments.
\end{abstract}


\setcounter{footnote}{0}
\setcounter{section}{0}

\section{Introduction}

\feii\ multiplet emission is a prominent feature in the ultraviolet (UV) to
optical spectra of most active galactic nuclei (AGNs),
including QSOs at redshifts as high as $\gtrsim 6$
(e.g., Barth et al. 2003; Freudling et al. 2003; Iwamuro et al. 2004;
Jiang et al. 2007; Kurk et al. 2007).
Its equivalent width (EW) varies significantly from object to object, ranging
from $>100$ \AA\ to undetectable [$<$5 \AA\ ($1\sigma$)\,]
for the \feii\,$\lambda4570$ blend in the 4434--4684 \AA\ region
(Boroson \& Green 1992).
\feii\ emission is an important probe of AGN physics.
For example, Boroson \& Green (1992) showed that the strength of
the \feii\,$\lambda4570$ emission relative to \oiii\ $\lambda5007$
is a dominant variable
in the first principal component (PC1 or Eigenvector 1) in their analysis
of the correlation matrix of QSO properties,
which is generally believed to be linked to
certain fundamental parameters of the accretion process
(e.g., the relative accretion rate,
often expressed as $L/L_{\rm Edd}$ or $L/M$;
Sulentic et al. 2000b; Boroson 2002).
\feii\ emission is a strong coolant of the broad-line region (BLR) of AGNs,
and thus provides useful information about the BLR from energy budget considerations
(Osterbrock \& Ferland 2006; see also Vestergaard \& Wilkes 2001, and references therein).
More importantly, \feii\ emission relative to that of $\alpha$-elements
(e.g., \mgii) can be considered to be an observational proxy
of the Fe/$\alpha$ abundance ratio.
If iron is mainly produced by Type 1a supernovae, which have long-lived progenitors,
this ratio would have a characteristic delay of $\sim$ 1 Gyr from the initial starburst
in the QSO host environment.
Thus, \feii\ emission in principle can serve as a cosmic clock
to constrain the age of QSOs and the epoch of
the first star formation in their host galaxies
(Hamann \& Ferland 1993; and cf., e.g., Matteucci \& Recchi 2001).

However, the ionization and excitation mechanisms of the \feii\ emission are very complex
due to the low ionization potential of the Fe$^0$ atom (7.9\,eV),
the low energy levels of the various excited states of Fe$^+$,
and particularly the complexity of its atomic structure
(see Baldwin et al. 2004; Collin \& Joly 2000 and references therein; Joly et al. 2008).
Thus, the conversion from \feii\ emission to iron abundance
is not straightforward (e.g., Baldwin et al. 2004; Verner et al. 2004).
Furthermore,
it is possible that \feii\ emission in AGNs arises from several different sites
with similarly suitable excitation conditions, including gaseous clouds
gravitationally bound in the BLR and narrow-line region (NLR),
(the base of) outflows, and the surface of the accretion disk
(see, e.g., Collin-Souffrin 1987;
Murray \& Chiang 1997; Matsuoka et al. 2007; Zhang et al. 2007).
This has already been hinted by the reverberation mapping observations
of \feii\ emission, particularly of its optical component,
since the cross-correlation function appears to be ill-defined,
being broad and flat-topped (Vestergaard \& Peterson 2005; Kuehn et al. 2008;
see also Kollatschny \& Welsh 2001; Wang et al. 2005).
Therefore, to use \feii\ emission as a proxy to estimate
Fe abundance and to constrain chemical evolution requires a better understanding of
its origin and emission mechanisms. In fact, it is well known that
the \feii/\mgii\ intensity ratios of QSOs at the same redshift have
a rather large scatter
(e.g., Dietrich et al. 2003; Iwamuro et al. 2004; Leighly \& Moore 2006).

It is also possible that the excitation mechanisms and
sites of \feii\ emission are well governed by certain
fundamental parameters (such as Eddington ratio, $\lratio$)%
\footnote{
The Eddington ratio, $\lratio$, is the ratio between the bolometric and
Eddington luminosities.
The Eddington luminosity (\ledd), by definition, is the luminosity at which the gravity
of the central source acting on an electron--proton pair (i.e. fully ionized gas)
is balanced by the radiation pressure due to electron Thomson scattering;
$\ledd = 4 \pi G c M \massp/ \st $,
where $G$, $c$, $M$, \massp, \st\
are the gravitational constant, speed of light, mass of the central source,
proton mass, and Thomson scattering cross-section, respectively.
In accretion-powered radiation systems, \lratio\ is often
referred to as the dimensionless accretion rate $\dot{m}$
(the accretion rate normalized by the Eddington accretion rate
$\dot{M}_{\rm Edd}$;
$\dot{m} \equiv \dot{M}/\dot{M}_{\rm Edd} = \eta c^2 \dot{M}/ \ledd$,
$\dot{M}$ being the mass accretion rate and $\eta$ the accretion efficiency),
as $\dot{m}$ is not a direct observable.  Yet, the two notations are different,
both in meaning and in scope of application. The
Eddington ratio applies to any radiation system,
whether accretion-powered or not.
Even in accretion-powered radiation systems like AGNs,
\lratio\ ($L$) is not equivalent to $\dot{m}$ ($\dot{M}$),
except in the simple thin accretion disk model of Shakura \& Sunyaev (1973)
[see, e.g., Merloni \& Heinz 2008].  This distinction should be kept
in mind.}
due to some self-regulation
mechanisms that maintain the normal dynamically quasi-steady
states of the gas surrounding the central engine of AGNs.
Recently, Dong et al. (2009a) found that
the EW of the \mgii\ $\lambda 2800$ emission line
does not depend intrinsically on AGN luminosity, broad-line width, or
BH mass, but is governed solely and strongly by \lratio.%
\footnote{This means that the Baldwin (1977) effect---the correlation between
emission-line EW and continuum luminosity---is a secondary effect of
the EW--\lratio\ relation (Dong et al. 2009a, 2009b).}
If this is also true for \feii\ emission, then, by accounting for the dependence on \lratio, it may be possible to calibrate the relationship between the
\feii/\mgii\ intensity ratio and the Fe/Mg abundance ratio, and thus
constrain chemical evolution in QSO environments.

Motivated by these considerations,
we carry out a systematic study of the strengths of UV and optical \feii\ emission,
by taking advantage of the unprecedented spectroscopic data from
the Sloan Digital Sky Survey (SDSS; York et al. 2000).
Using the identification and measurement of \feii\ emission lines
from V\'eron-Cetty et al. (2004),
we are also able to study narrow-line \feii\ emission systematically  for the first time.
The paper is organized as follows.
In Section 2 we describe the selection criteria of the sample,
the spectral fitting and measurements, investigation of \feii\ strength with respect
to continuum and other emission lines, and correlation and regression analyses.
Section 3 presents the results and our discussion.
Section 4 gives conclusions and implications.
Throughout this paper, we use a cosmology with
$H_{\rm 0}$ = 70 km\,s$^{-1}$\,Mpc$^{-1}$, $\Omega_{\rm m}$ = 0.3, and
$\Omega_{\rm \Lambda}$ = 0.7.

\section{Data Analysis}
\subsection{Sample and Spectral Fitting}

\subsubsection{Sample Construction}
We first construct a homogenous sample of Seyfert 1s and QSOs
(namely type 1 AGNs)
from the spectral data set of the SDSS Fourth Data Release
(Adelman-McCarthy et al. 2006), according to the following criteria:
(1) redshift $z \leq 0.8$;
(2) median signal-to-noise ratio (S/N) $\geq 10$ pixel$^{-1}$ in the
optical \feii\ and \hb\ region (4400--5400 \AA);
(3) weak stellar absorption features, such that the rest-frame EWs of \caii\,K
(3934 \AA),
\caii\,H~+~H$\epsilon$ (3970 \AA), and H$\delta$ (4102 \AA) absorption features
are undetected at $< 2\,\sigma$ significance.
The redshift cut is set to ensure that \hb\ and the \feii\,$\lambda4570$
multiplets are present in the SDSS bandpass.
The S/N criterion allows
proper placement of local continua and the accurate measurement of
emission lines (especially broad and narrow \feii\ emission).
The third criterion ensures that the AGN luminosity and
emission-line EWs suffer minimally from
contamination by host galaxy starlight.
Normally, the \caii\,K absorption feature alone can effectively gauge the
level of starlight contamination; however, in AGN-dominated spectra,
the measurement of this feature in practice can be affected by
nearby emission lines in the continuum windows.  So we visually inspect the
spectra that have \caii\,K absorption detected at $\geq 2\,\sigma$
but no \caii\,H~+~H$\epsilon$ or H$\delta$ absorption features detected
at $\geq 1\,\sigma$ significance. A small fraction ($\sim 10\%$) of the objects
are retained in this way.  To quantify the level of galaxy contamination
imposed by our selection process, we simulated artificial spectra by
combining different proportions of template spectra of pure AGNs
(high-luminosity QSOs) and pure starlight (absorption-line or weak
star-forming galaxies).  As described in greater detail in the Appendix,
our selection criterion corresponds to a galaxy contribution of
$\lesssim 10$\% around 4200~\AA.
Compared to other sources of errors (see Section 2.1.2), this level of
starlight contamination has little effect on the emission-line measurements;
it contributes at most 0.002~dex (0.5\%) to the variance of emission-line EWs.

After removing duplications and sources with 
excessive bad pixels in the \hb\ or \feii\,$\lambda 4570$ region,
our final sample consists of 4178 type~1 AGNs
(hereinafter the full type 1 sample).
Of these, 2092 have redshifts $z \gtrsim 0.45$ and
median S/N $\gtrsim 10$ pixel$^{-1}$ in the UV \feii\ region (2200--3090 \AA);
this UV subsample will be used to examine the behavior of UV \feii\
and \mgii\ $\lambda 2800$.

\subsubsection{Spectral Fitting and Measurements}
We use the measured parameters from the Valued-added ExtraGAlactic Catalog
developed and maintained by the Center for Astrophysics at the University of
Science and
Technology of China (USTC--VEGAC; X.-B.~Dong et al., in preparation;
J.-G.~Wang et al., in preparation).
Details of the spectral fitting in the optical and near-UV regions have been
described in Dong et al. (2008) and Wang et al. (2009), respectively.
Here we present a brief description of the continuum and emission-line fits.
The fits are based on the MPFIT package (Markwardt 2009), which performs
\chisq-minimization using the Levenberg--Marquardt technique.
We corrected the spectra for Galactic extinction using the extinction
map of Schlegel et al.\ (1998) and the reddening curve of Fitzpatrick (1999).

The optical \feii\ is modeled with two
separate sets of templates in \emph{analytical} forms,%
\footnote{The implementation of the two template functions in Interactive
Data Language (IDL) is available at
http://staff.ustc.edu.cn/\~{ }xbdong/Data\_Release/FeII/Template/ \,.}
one for the broad-line system and the other for the narrow-line system.
These two sets of templates are constructed from measurements of I\,Zw\,1 by
V\'eron-Cetty et al. (2004), as listed in their Tables~A1 and A2;
see Dong et al. (2008) for details.
Within each system, the respective set of \feii\ lines are assumed to
have no relative velocity shifts and the same relative strengths as those in I~Zw~1.
The UV \feii\ is modeled with the tabulated semi-empirical template for
I~Zw~1 generated by Tsuzuki et al. (2006).
In the wavelength region covered by \mgii\ emission,
they employed a semi-empirical iteration procedure to build the template.
They first generated a theoretical \feii\ model spectrum with
the photoionization code CLOUDY (Ferland et al. 1998) and subtracted it
from the observed spectrum of I\,Zw\,1 around \mgii.
Then the \mgii\ doublet lines were fit assuming each line has the same profile as \ha.
And finally they obtained the \feii\ template underneath \mgii\
by subtracting the \mgii\ fit from the observed spectrum.
The separation of narrow and broad lines was not taken into account
for the UV \feii\ template, but narrow lines in the UV are generally weak
(e.g., Laor et al. 1994).

Each line in the \mgii\ doublet is fitted with a truncated five-parameter
Gauss-Hermite series (Wang et al. 2009; see also Salviander et al. 2007).
As the profile of broad \hb\ is sometimes rather complex,
it is fitted with as many Gaussians as statistically justified;
we do not ascribe any particular physical significance to
the individual Gaussian components (Dong et al. 2005).
We assume that the broad \feii\ lines have the same profile as broad \hb\
(see Boroson \& Green 1992; Landt et al. 2008; cf. Hu et al. 2008b), but
we leave their relative velocity as a free parameter.
The fluxes of both broad and narrow \feii\ emission are quite insensitive
to the exact line width (Vestergaard \& Peterson 2005; Landt et al. 2008)
and profile shape assumed for the broad component, which is verified at
the end of this section.
All narrow emission lines are fitted with a single Gaussian, except for the
\oiii\ $\lambda \lambda$4959, 5007 doublet lines, each of which is
modeled with two Gaussians, one accounting for the line core and the other
for a possible blue wing seen in many objects.
The redshift and width of narrow-line \feii\ are set as free parameters.

The presence of optical narrow \feii\ emission lines in AGNs has been long
overlooked.  Previously, narrow \feii\ emission has been studied in only a few
objects (in the near-UV: Vestergaard \& Wilkes 2001; Wang et al. 2008; in the
optical: V\'eron-Cetty et al. 2004, 2007; Bruhweiler \& Verner 2008;
cf. Section 4.6 of Hu et al. 2008b).  A companion paper (Dong et al. 2010),
drawn directly from the current sample and analysis, presents the first
systematic study of the prevalence of optical narrow \feii\ emission lines in
type 1 AGNs and their non-detection in type 2 AGNs.  Whether there is a physically
distinct boundary
between the NLR and BLR is still debated (cf. Laor 2007), particularly
considering the likely presence of a region intermediate between them
(e.g., the intermediate-line region, Brotherton et al. 1994;
the inner NLR, Nagao et al. 2003; see Zhang et al. 2009 for a concise review),
and the fact that both the BLR and NLR may be highly stratified and multi-zoned.
In this paper, we follow V\'eron-Cetty et al. (2004), who divided the
\feii\ spectrum phenomenologically into two main kinematic sub-systems: a
broad (L1; FWHM $\approx 1100$ \kms)
component associated with other prominent broad emission lines, and
a narrow (N3; FWHM $\approx 280$ \kms) component, consisting of both permitted
and forbidden transitions,
associated with other low-excitation narrow lines.  In I\,Zw\,1, there are two
additional narrow-line systems, N1 and N2, that appear only in high-excitation
lines (not in \feii); N1 and N2 are relatively broad and blueshifted
(by 1450 and 500 \kms, respectively), whereas N3 is almost at the systemic
redshift of the host galaxy.  The broad-line system L1 is also blueshifted by
150--200 \kms\ in both the optical and UV bands (Laor et al. 1997;
V\'eron-Cetty et al. 2004), which distinguishes it from the narrow-line system
N3 and made possible the unambiguous identification of narrow \feii\ lines
(P. V\'eron 2009, private communications).

Figure~\ref{fig:optfitting} shows the optical spectrum and spectral
decomposition of a representative object in our sample
(SDSS\,J092801.30+491817.3).  In this object, the optical \feii\ lines
have a width (corrected for the instrumental broadening of FWHM = 130 \kms) of
FWHM = 1300 \kms\ for the broad component and FWHM = 250 \kms\ for the
narrow component.  Note that the individual narrow \feii\ lines marked in the
figure (particularly the forbidden lines) are
sharp and distinct from nearby broad \feii\ lines,
which indicates that our identification and decomposition
of the narrow component is robust, not residuals from poor broad \feii\
subtraction due to mismatch of the broad \feii\ model.
Strong narrow \feii\ emission is found in AGNs of diverse types:
broad-line and narrow-line Seyfert 1s, QSOs with and without broad
absorption lines,
radio-loud and radio-quiet systems.
Figure~\ref{fig:demoobjs} displays a sample of spectra in which
strong narrow \feii\ emission is seen; the object shown in the bottom panel is
the low-ionization broad absorption-line QSO FBQS\,J152350.4+391405, which is a powerful
($L_{\rm 5GHz}= 4.0 \times 10^{31}$ erg\,s$^{-1}$\,Hz$^{-1}$), variable radio
source (Becker et al. 2000).  An example of spectral fitting for the
UV subsample is shown in Figure~\ref{fig:uvfitting}.
The detailed decompositions of the line profiles of \hb\ and \mgii\
have been demonstrated in Dong et al. (2008; their Figure~2) and
Wang et al. (2009; their Figure~1), respectively.

The emission-line fluxes are measured from
the best-fit models of the line profiles.
The EWs are calculated in the rest frame from the best-fit models of
both the emission lines and their underlying local continuum,
by integrating the line profile with respect to the continuum level
pixel by pixel.
The fluxes and EWs of narrow and broad optical \feii\,$\lambda4570$ emission
are integrated in the wavelength range 4434--4684 \AA,
and those for UV \feii\ are integrated in the range 2200--3090 \AA.
For all measured emission-line fluxes and EWs, we regard the values as reliable
detections when they have greater than $2\,\sigma$ significance; otherwise,
we adopt the value plus the $2\,\sigma$ error as an upper limit (see
Section~2.1.3 for the estimation of measurement errors).
Using $2\,\sigma$ significance instead of the more commonly used
$3\,\sigma$ one is a trade-off we have made in order to obtain a sample
sufficiently large to explore parameter space,
particularly for narrow \feii\ (see Figure 11).  We have verified that
none of our main conclusions are affected by this particular choice.
The continuum and
emission-line parameters for the optical and UV subsamples are listed in
Tables 1 and 2, respectively; we make available online all the detailed
fitting results.\footnote{Available at
http://staff.ustc.edu.cn/\~{ }xbdong/Data\_Release/ell\_effect/,
together with auxiliary code to explain and demonstrate the fitting
and the parameters.}

Lastly, to test the effect of our adopted 
broad \feii\ line profile
on the flux measurements of both broad and narrow \feii\ emission,
we experimented with an alternative scheme in which a single Lorentzian with adjustable
width is used to model the profile of the individual broad \feii\ lines.
This choice is inspired by the fact that the broad \feii\ lines in I~Zw~1 are
well described by a Lorentzian profile (V\'eron-Cetty et al. 2004).
This alternate scheme yields broad \feii\ line widths that are on average
only 0.3 times that of broad \hb\ with a standard deviation of 0.2~dex (46\%),
but the line fluxes---the main
focus of this study---are statistically unchanged.  We find that
the fluxes of broad $\feii\,\lambda4570$ emission
(hereinafter $\feii^{\rm B}\,\lambda4570$) are similar to those of our default
scheme within 0.3~dex (70\%)
for 97\% (4055/4178) of the sample, while
the fluxes of narrow $\feii\,\lambda4570$ (hereinafter $\feii^{\rm N}\,\lambda4570$),
agree to within 0.3~dex for 88\% of the objects.  This is illustrated in
Figure~\ref{fig:testschemes}, where, for clarity, we only plot the objects
with fluxes detected at $>3\,\sigma$ significance by both schemes.
As already reported in Vestergaard \& Peterson (2005, their Section 2.3.4)
and Landt et al. (2008, their Section 4.4), the broad \feii\ multiplets are so
highly blended that their summed overall profile mainly depends on the
relative strengths of the multiplets rather than on their velocity widths.
According to their experiments, changing the width of the broad \feii\ template
(constructed from I\,Zw\,1, which has FWHM $= 1100$ \kms) by as much as
several thousand \kms\ has only a minor effect on the resulting line fluxes,
especially for the cases with larger line widths. 
Thus, we are confident that
the fluxes of both the broad and narrow \feii\ emission, integrated over a
large wavelength range, are very insensitive to the exact profile shape or
width assumed for the individual broad lines.

\subsubsection{Estimation of Parameter Uncertainties} 

The errors on the fitted parameters provided by MPFIT only account for formal
statistical uncertainties and likely underestimate the true uncertainties.
They do not include potential systematic uncertainties introduced by, for
example, line deblending or pseudocontinuum subtraction (see Marziani et al.
2003 for a detailed discussion).

We estimate the measurement uncertainties due to line deblending using a
bootstrap method (Dong et al. 2008; Wang et al. 2009), as follows.  We
generate 500 spectra by randomly combining the scaled, model emission lines of
one object (denoted as ``A'') to the emission line-subtracted spectrum of
another object (denoted as ``B''). In order to minimize changes in S/N within
the emission-line spectral regions in the simulated spectra, the emission-line
model of object ``A'' is scaled in such a way that it has the same flux for
the line in question as in object ``B.'' Then, we fit the simulated spectra
following the same procedure as described in Section 2.1.2.  For each parameter,
we consider the error typical of our sample to be the standard deviation of
the relative difference between the input and the recovered parameter values.
These relative differences turn out to be normally distributed for each of the
parameters concerned.  The estimated typical $1\,\sigma$ relative errors
are 0.043~dex (10\%), 0.035~dex (8\%), 0.043~dex (10\%), 0.052~dex (12\%),
0.048~dex (11\%), and 0.074~dex (17\%), respectively, for the fluxes of
broad \mgii, broad \hb, \oiii\,$\lambda 5007$, UV \feii,
broad \feii\,$\lambda 4570$, and narrow \feii\,$\lambda 4570$;
0.087~dex (20\%) and 0.065~dex (15\%), respectively,
for the FWHM of broad \mgii\ and \hb;
and 0.035~dex (8\%) and 0.022~dex (5\%), respectively,
for the slope and normalization of the local continua.
The errors on EWs are calculated using standard propagation of errors.
In the analysis of Sections 2.2 and 2.3, we generally adopt the errors based
on the bootstrap method, with the exception of a minority of objects in which
the bootstrap errors are actually smaller than the formal MPFIT errors, in which
case we adopt the latter.

The uncertainties due to pseudocontinuum subtraction are harder to estimate.
Two factors come into play.  The first is due to the choice of \feii\ template.
Although we have adopted the latest improvements to the \feii\ template
(V\'eron-Cetty et al. 2004; Tsuzuki et al. 2006), the fact remains that
essentially all templates used in this field, including ours, are derived
ultimately from observations of a single AGN, namely I~Zw~1.  Our choice of
using these templates is motivated entirely by pragmatism: empirically,
they seem to work, and they are the best we have at the moment.  Unfortunately,
there is currently no practical way to quantify the uncertainties that
might be introduced by this restriction.

For a given choice of \feii\ template, additional uncertainties arise from the
fitting procedure, since we must assume a profile for the broad \feii\ lines,
which are too highly blended to be determined independently.  Our default
fitting scheme---motivated by previous studies---assumes that the broad
components of \feii\ and \hb\ have exactly the same profile.  In detail, of
course, this cannot be strictly true.  To estimate the likely impact of
this assumption, we refit the spectra assuming that \feii\ has a Lorentzian
profile (Section 2.1.2 and Figure 4).  The standard deviations of the
differences in flux between our default scheme and the Lorentzian scheme are
0.09~dex (21\%) and 0.12~dex (28\%)
for broad and narrow \feii\,$\lambda4570$, respectively.
We suspect that uncertainties of roughly this magnitude ($\sim 0.1$ dex)
can potentially affect the fluxes of all the \feii\ lines, as well as those of
\mgii\ and \oiii, which are strongly affected by \feii\ contamination.  This
additional uncertainty was added to the error budget of the fluxes of these
lines.  We do not consider this error contribution to the fluxes of other
narrow lines or of broad \hb, the bulk of which is not severely affected by
\feii\ emission.

The continuum luminosities employed in our analysis are affected not only by
\feii\ subtraction, but also, to some degree, by host galaxy contamination,
despite our efforts to mitigate it (see the Appendix).  The 3000 \AA\ continuum
is additionally influenced by our treatment of the Balmer continuum
(Wang et al. 2009).  Taking all of these factors into consideration, we
estimate that the continuum luminosities at 3000 \AA\ and 5000 \AA\ incur
an uncertainty of $\sim 0.05$ dex (12\%) on top of that derived from the
bootstrapping method.

Typical error bars, which represent the quadrature sum of the uncertainties
described above, for the parameters used in our analysis are shown in Figures
5--12.  However, because these estimates are only approximate, and it is
nearly impossible to derive rigorous errors for every individual object, we
will adopt only the bootstrapping errors in the regression
analysis below.  We will not attempt to estimate the intrinsic scatter of
the relations presented in this paper.

\subsection{The Strength of \feii\ Emission}

We calculate the fluxes and rest-frame EWs of strong emission lines and
investigate the distributions of the EWs of various \feii\ emissions and
their relative strengths with respect to other lines.
We have already presented in Dong et al. (2010, their Figure~3)
the distributions of the EWs of $\feii^{\rm N} \lambda 4570$ and
$\feii^{\rm B} \lambda4570$; the two quantities do not correlate very strongly%
\footnote{Throughout the paper, we regard a correlation as statistically
significant when the probability of obtaining the null hypothesis that the
correlation is not present (\pnull) is less than 1 per cent,
and regard one as strong when it is significant and has a Spearman correlation
coefficient $\rs \gtrsim 0.5$.
Since the correlation test used in our analysis does not handle
the attenuation of correlation strength caused by measurement error,
the correlation strengths reported here are likely to
be weaker than their intrinsic values; this should be particularly true for
the variables whose dynamical range is relatively small with respect to the
measurement error.  This effect, however, will not nullify
the already significant/strong correlations presented in this paper.
}
(Spearman coefficient $\rs =0.45$, hereinafter in this subsection accounting
for upper limits; see Section 2.1.2 for the definition of upper limits),
suggesting that the narrow component is
not an artifact of measurement uncertainty associated with
the deblending of the broad component.
The intensity ratios of narrow to broad \feii\,$\lambda4570$,
for the 2502 objects in which both components are detected at
$>3\,\sigma$ significance (see Dong et al. 2010),
vary by 2 orders of magnitude, ranging from
$\sim$0.005 to 0.5, with a mean of $-1.15$~dex (equivalently 0.07 in linear scale;
computed in log-scale, hereinafter for the quantities in this subsection) and
a standard deviation of 0.30~dex.
The distribution of the EWs of $\feii^{\rm B} \lambda4570$ and UV \feii\
emission is displayed in Figure~\ref{fig:ewb4570vsuvfeii}
for the 2092 objects in the UV subsample.
From the EW--EW distribution, the \feii\ emissions in the two bands do not
appear to be strongly correlated ($\rs=0.29$).
Yet, this comparison might be complicated by variations in the continuum
shape, since the underlying continuum of the two bands used for the EW
calculation spans across a considerable wavelength range.  In
Figure~\ref{fig:fluxb4570vsuvfeii} (left panel), we plot
instead the distribution of the fluxes of the two emission blends, which now
show a fairly strong correlation ($\rs=0.68$).  This is also reflected in the
histogram of their flux ratios (right panel), for 2076 of the 2092 objects
wherein both blends
are reliably detected (at $\geq 2\,\sigma$ significance, see Section 2.1.2),
which clusters around a mean of $-0.96$~dex (0.11) with a
standard deviation of only 0.25~dex (see also Sameshima et al. 2011).
A similar result holds for the
relationship between UV \feii\ and \mgii\ ($\rs=0.69$;
Figure~\ref{fig:uvfeiivsbmgii});%
\footnote{In Figures 7, 8 and 9, we revert back
to plotting EWs (instead of fluxes) for the distribution diagrams because the
features on both axes are very close in wavelength, and hence their comparison
is not distorted by possible variations in the shape of the continuum.}
the ratios of UV \feii\ to \mgii\ flux peak at 0.67~dex (4.70) with a standard deviation
of 0.21~dex.  $\feii^{\rm B} \lambda4570$ and broad \hb\ are
less strongly correlated ($\rs = 0.55$; Figure~8); their flux ratios have a mean
of $-0.33$~dex (0.47) and a standard deviation of 0.30~dex.  The strength of
$\feii^{\rm N} \lambda4570$ does not correlate at all with that of
\oiii\,$\lambda5007$ (Spearman chance probability
$\pnull = 0.2$ for their flux--flux relationship;
see also Figure~\ref{fig:n4570vsoiii}a for their EW--EW distribution).
In order to check if the measurement of narrow \feii\ is biased by
the narrow \feii\ template we used,
we further calculate the EWs of two other narrow \feii\ features directly from the
residual spectra (after the continuum, broad \feii,
and other emission lines, except narrow \feii, are subtracted), namely
\feii\,$\lambda4925$ (integrated over the vacuum wavelength range
4918--4938~\AA, which is dominated by \feii\,42 $\lambda4923$ and
\feii]\,$\lambda4928$) and \feii\,49 $\lambda5234$.
These two features are relatively distinct from nearby broad \feii\ features.
The distributions of their EWs are also displayed in
Figure~\ref{fig:n4570vsoiii} (panels b and c); they also show no strong correlation
with \oiii\ at all.

\subsection{Correlation and Regression Analysis}

We investigate the correlations of the EWs and intensity ratios
of narrow and broad \feii, broad \mgii\ and \hb, and \oiii\ $\lambda 5007$
with broad-line FWHM, continuum luminosity $L_{5100} \equiv $\lfive, \mbh, and \lratio.
We calculate the black hole (BH) masses based on \hb\ using
the formalism presented in Wang et al. (2009, their Eqn.~11).
This formalism was calibrated with recently updated reverberation
mapping-based masses and assuming that the BLR radius scales with luminosity as
$R \propto L^{0.5}$ (Bentz et al.
2009).  The Eddington ratios are estimated assuming a bolometric
luminosity correction $\lbol \approx 9 \,$\lfive\ (Elvis et al. 1994; Kaspi et al.
2000).  The mean and standard deviation (computed in log-scale) of the key
variables of the sample are as follows: FWHM of
broad \hb, 3.56~dex (equivalently 3600 \kms\ in linear scale) and 0.22~dex;
\lfive, 44.60~dex ($4.0 \times 10^{44}$ \lum) and 0.40~dex;
\mbh, 8.30~dex ($2.0 \times 10^{8}$~\msun) and 0.35~dex;
\lratio, $-0.85$~dex (0.14) and 0.28~dex.
We assume the $1\,\sigma$ measurement errors for \mbh\ and \lratio\
to be 0.3~dex (70\%; Wang et al. 2009).

We perform the bivariate correlation tests using the generalized Spearman rank method
implemented in the ASURV package (Isobe et al. 1986).
This method tests for not only a linear relation but a monotonic one,
and it can handle censored data in both independent and dependent variables.
The correlation results are summarized in Table 3, where
we report the Spearman coefficient (\rs) and the probability
($P_{\rm null}$) that a correlation is not present.
Several striking features emerge from the correlation analysis:
\begin{itemize}
    \item
    The strongest correlations for the EWs and intensity ratios of
    all emission lines are with either \lratio\ or broad-line FWHM.
    For a particular emission-line EW or intensity ratio,
    the correlations with FWHM and with \lratio\ are almost equally strong,
    and both are much stronger than those with \mbh\ or $L$.
    The correlation with $L$ is generally the weakest.

    \item
    The EW of UV \feii\ has \emph{no} significant correlation with \lratio\
    (\pnull = 0.04), or with the other three quantities, but
    the EWs of optical \feii, both broad and narrow, show moderate to strong, positive
    correlations with \lratio\ ($\rs = 0.67$ for narrow \feii\
    and $\rs$ = 0.40 for broad \feii, both with $\pnull \ll 10^{-15}$).

    \item
    The intensity ratios of \feii---both narrow and broad,
    both in the UV and in the optical---to \mgii\ $\lambda2800$ correlate
    strongly and positively with \lratio.  Among these, the
    strongest correlation arises from the narrow component of
    \feii\ (\rs = 0.70).
    Interestingly, the intensity ratios of optical \feii\ (narrow and broad)
    to \mgii\ correlate more strongly with \lratio\ than do the
    EWs of these lines.

\end{itemize}

We must note that because the SDSS spectroscopic survey is magnitude-limited,
broad-line FWHM, $L_{5100}$, \mbh, and \lratio\ show apparent
correlations with one another.
The apparent (likely not intrinsic) correlation between \mbh\ and \lratio\
is further enhanced by the correlation of their measurement uncertainties,
because both \mbh\ and \lratio\ are constructed from
\hb\ FWHM and $L_{5100}$.
The Spearman correlation coefficients of \lratio\
with \hb\ FWHM, $L_{5100}$, and \mbh\ are \rs\ = $-0.70$, $0.52$, and $-0.19$,
respectively, for the full sample; for the UV subsample, they are
\rs\ = $-0.82$, $0.48$, and $-0.40$, respectively.
In light of the serious inter-dependence among these four quantities,
the correlations of emission-line EWs and intensity ratios with $L_{5100}$ and \mbh\
are probably only a secondary effect of the stronger (thus presumably intrinsic)
correlation with \lratio\ (or broad-line FWHM).
To test this possibility, we perform a partial correlation analysis
using the generalized partial Spearman rank method
(Kendall \& Stuart 1979; Macklin 1982).
The partial correlation results are summarized in Table~\ref{tab-partialcorr}.
Because of the strong inter-dependence among the four physical variables,
unfortunately, even partial correlation tests cannot definitively
discriminate which variable is the primary driver.  However, several trends
do stand out clearly:


\begin{itemize}

\item
First, for all the EWs and intensity ratios that still have significant correlations
with \mbh\ or $L_{5100}$ controlling for \lratio\ ($\pnull \lesssim 10^{-3}$),
their correlations with \lratio\ controlling for \mbh\ or $L_{5100}$
are also significant ($\pnull \ll 10^{-15}$).
This is just as expected in light of the bivariate correlations,
which are weaker with \mbh\ and much weaker with $L_{5100}$ than with \lratio.

\item
Second, for almost all the EWs and intensity ratios (except the EW of broad
\hb\ and the ratio of broad \feii\,$\lambda4570$ to UV \feii) that have
significant correlations with FWHM controlling for \lratio\ ($\pnull < 10^{-3}$),
their correlations
with \lratio\ controlling for FWHM are also significant
($\pnull \leq 10^{-10}$).

\item
Third and most important,
for some key emission-line quantities, namely
the EWs of broad \feii\,$\lambda4570$, \mgii\,$\lambda2800$,
and \oiii\,$\lambda5007$,
the ratio of broad \feii\,$\lambda4570$ to \oiii\,$\lambda5007$
(the dominating variable of the PC1 of Boroson \& Green 1992),
and the ratio of UV \feii\ to \mgii\,$\lambda2800$
(the common proxy for abundance ratio Fe/$\alpha$),
their correlations are very significant with \lratio\ controlling for
$L_{5100}$, \mbh, or FWHM ($\pnull \ll 10^{-15}$),
but are much less significant (or not significant at all)
with $L_{5100}$, \mbh, and even FWHM controlling for \lratio\
($\pnull > 10^{-8}$ and $\rs \leq 0.1$).
The best example is the ratio of broad \feii\,$\lambda4570$ to
\oiii\,$\lambda5007$, which shows no correlation with $L$, \mbh, or
FWHM at all ($\pnull > 0.1$) controlling for \lratio.
Another example is the EW of \mgii\,$\lambda2800$,
which was investigated thoroughly in Dong et al. (2009a).
This suggests that, \emph{at least for these important emission-line
EWs and intensity ratios, their apparent correlations with broad-line FWHM,
continuum luminosity, and \mbh\ are
mainly a secondary effect of their relationship with \lratio.
\lratio\ is the principal, if not sole, physical driver.}

\end{itemize}

Regarding other emission-line EWs and intensity ratios (e.g., the ratio of
broad \feii\,$\lambda4570$ to UV\,\feii, the ratio of narrow
\feii\,$\lambda4570$ to \oiii\,$\lambda5007$;
cf. Kova{\v c}evi{\'c} et al. 2010; Sameshima et al. 2011),
it is hard to tell from the statistical tests
whether broad-line FWHM or \lratio\ is the primary driver.
It is not surprising that their
correlations with FWHM are very close to, or even slightly stronger than, that
with \lratio, since \lratio\ depends strongly on FWHM, by construction.
It is possible that
the intrinsic, primary driver is indeed \lratio, but that the statistical tests are
obscured by systematic uncertainties plaguing the estimated values of
\lratio.  One effect is the large uncertainties in virial BH masses, which can
be a factor of 4 ($1\sigma$) statistically, and perhaps as large as an order
of magnitude for individual estimates (Vestergaard \& Peterson 2006; Wang et
al. 2009). Another uncertainty comes from the bolometric correction assumed
for $L_{5100}$, which is definitely an oversimplification in light of the diverse
spectral energy distributions of AGNs (Ho 2008; Vasudevan \& Fabian 2009;
Grupe et al. 2010).

Figure~\ref{fig:blewcorrs} examines the strength of broad \feii\ emission
and its dependence on three AGN physical parameters, $L_{5100}$, \mbh, and
$\lbol/\ledd$.  The same is repeated in Figure~\ref{fig:nlewcorrs} for
narrow \feii.  Lastly, Figure~\ref{fig:femgell} explores variations of the
ratios of broad and narrow \feii\ to \mgii\ with respect to \lratio.  The
strong correlations with \lratio\ are striking considering the narrow range of \lratio\ in
our UV subsample ($1\,\sigma = 0.26$~dex for a log-normal distribution) and the
possible systematic errors in \lratio, as discussed above.
It is particularly noteworthy that intensity ratios of narrow and broad
optical \feii\ to \mgii\ correlate more strongly with \lratio\ than do
the EWs of these lines (see Table \ref{corrtab}).
We performed linear regressions (in log--log scale)
using the LINMIX code of Kelly (2007).
This method adopts a Bayesian approach and accounts for measurement
errors, censoring, and intrinsic scatter.
The results are as follows:
\begin{eqnarray}
  \log \frac{{\rm Fe\,II}^{\rm N} \lambda4570}{\rm Mg\,II} & ~=~ &
  (0.40\pm 0.11)  ~+~  (2.46 \pm 0.15) \log \lbol/\ledd \\
  \log \frac{{\rm Fe\,II}^{\rm B} \lambda4570}{\rm Mg\,II} & ~=~ &
  (0.74\pm 0.08)   ~+~  (1.23 \pm 0.10) \log \lbol/\ledd \\
  \log \frac{\rm Fe\,II~UV}{\rm Mg\,II} & ~=~ &
  (1.21\pm 0.07)   ~+~  (0.63 \pm 0.09) \log \lbol/\ledd ~ .
\end{eqnarray}
The intrinsic standard deviations of these relations (red lines in Figure 12),
corrected for measurement errors as given by LINMIX,
are 0.05~dex, 0.18~dex, and 0.14~dex, respectively.

%
To check for possible effects of BH mass estimation on our results, we
reexamine the above correlation tests with \mbh\ calculated using several
other commonly used virial mass formalisms based on broad \hb\ and/or \mgii\
(McLure \& Dunlop 2004; Collin et al. 2006; Vestergaard \& Peterson 2006;
Vestergaard \& Osmer 2009).  The alternate masses give similar results to
those listed in Table 3 (see also Table 1 of Dong et al. 2009a).
This is mainly because the dynamical range on \mbh\ covered by our sample
is not very large ($\sim 2.3$ dex for the entire sample and $\sim 1.5$ dex for
the UV subsample, centered at $\sim 2\times10^8$ \msun), and in this range the
various formalisms based on single-epoch
\hb\ or \mgii\ have only subtle differences from one another (Wang et al. 2009).


We also reevaluate the correlations using the continuum luminosity in the UV
and the Eddington ratio estimated from it,
based on the 2092 sources in the UV subsample.
The UV luminosity is calculated from the best-fit continuum at 2500~\AA,
$L_{2500} \equiv \lambda L_{\lambda}(2500~$\AA), and the
Eddington ratio is estimated assuming a bolometric
correction $\lbol \approx 6.3 \, L_{2500}$ (Elvis et al. 1994).
The correlation results are very similar to those listed in Table \ref{corrtab}
(see also Table 1 of Dong et al. 2009a).
For instance, the correlations of the EW of broad \feii\,$\lambda4570$
with $L_{2500}$ and with the corresponding \lratio\ (derived from $L_{2500}$)
have $\rs = 0.03$ and 0.39, and $\pnull = 0.1$ and $< 10^{-15}$, respectively;
for the correlations with the EW of narrow \feii\,$\lambda4570$,
$\rs = 0.03$ and 0.64, and $\pnull = 0.1$ and $< 10^{-15}$, respectively.
These results confirm our conclusion that the EWs of narrow and broad optical
\feii\ significantly correlate with \lratio\ but not with AGN luminosity
intrinsically.  Similarly, the correlations of the EW of UV \feii\
with $L_{2500}$ and the corresponding \lratio\ are still very weak,
having $\rs = -0.10$ and $-0.11$,
although their significance has increased, to
$\pnull = 2 \times 10^{-6}$ and $< 10^{-15}$, respectively.
The enhanced significance is probably caused by the fact that
the EW of UV \feii\ itself depends on the UV continuum by definition.
In any event, the EW of UV \feii\ correlates much more weakly, if at all,
with \lratio\ than is the case for the EWs of optical \feii.
The lack of correlation between UV \feii\ and the optical or near-UV
continuum is not very surprising, because in the photoionization picture UV
\feii\ is powered by the continuum at shorter wavelengths.

\section{Results and Discussions}

\subsection{\lratio\ Controls the Strength of Narrow and Broad Optical \feii}

As shown above, a general trend echoed throughout our analysis is that
the emission-line EWs and intensity ratios
correlate more strongly with \lratio\ than with $L$ or \mbh.
This paper focuses on the behavior of narrow and
broad \feii\ emission, particularly in the optical.  We highlight three points.

\begin{enumerate}
\item
First, it is quite unexpected that narrow, rather than broad, \feii\ emission
correlates more strongly with \lratio,
as the gas emitting broad \feii\ should be closer to,
and thus more tightly linked with, the central engine
than that associated with narrow \feii.
One possible explanation is that the origin of narrow \feii\ is more
homogeneous than that of broad \feii\ (see Section 3.2).

\item
Second, the EWs of both UV and optical broad \feii\ vary significantly
from object to object, with a similar amplitude of about 1.5~dex
(see Figure \ref{fig:ewb4570vsuvfeii});
yet, unlike narrow and broad optical \feii\ emission,
the EW of UV \feii\ has no correlation at all with \lratio.

\item
Third and probably most important, as mentioned in Section 2.3 and
shown in Figure~\ref{fig:femgell}, the ratios of \feii\ to \mgii\
correlate strongly with \lratio, but probably \emph{not} intrinsically with
broad-line FWHM, $L$, or \mbh.
This is also the case for the EWs of broad optical \feii, \mgii, and \oiii, as well as the
ratio between broad optical \feii\ and \oiii\,$\lambda5007$,
the dominant variable of the PC1 of Boroson \& Green (1992).
We will discuss this issue further below. 
\end{enumerate}


A strong, negative correlation between the EW of \civ\ $\lambda 1549$ and
\lratio\ has been noted by Baskin \& Laor (2004) and Bachev et al. (2004).
Both groups suggested that \lratio\ is the fundamental driver of the Baldwin
effect---a well-known inverse correlation between emission-line EWs and AGN
luminosity (Baldwin 1977).  The findings in this paper expand this picture:
although the gas environment in the line-emitting region of AGNs may be
complex and chaotic, \emph{the strength of 
several important emission lines (\civ\ $\lambda 1549$, \mgii\ $\lambda 2800$,
and optical Fe\,{\footnotesize II}) are governed by \lratio}.
At face value, from a statistical point of view some correlations
are equally strong with broad-line FWHM (e.g., the EW of broad \hb; but
definitely not for broad \mgii, see Dong et al. 2009a; cf. Boroson et al.
1985; Wang et al. 1996).  However, there is no obvious physical process
closely related to the line width that can easily explain the above statistical
trends.  Instead, we propose a unified picture governed by \lratio.

The high-ionization line \civ\ $\lambda 1549$ is produced by ionizing photons
above 47.85 eV.  \mgii\ $\lambda 2800$, a low-ionization line, is
collisionally excited from Mg$^+$ ions, which are produced by
photons above 7.65 eV and destroyed by photons above 15.04 eV.  Moreover, the
Mg$^+$ ion can be destroyed by the diffuse Balmer radiation field,
and \mgii\,$\lambda2800$,
being an optically thick line, can be scattered and absorbed by excited \hi\
atoms.  As with \mgii, \feii\ is produced by photons above 7.9 eV and
destroyed by photons above 16.2 eV.  However, the optical \feii\ lines, being
completely optically thin, do not suffer at all from absorption by excited \hi\ atoms.%
\footnote{See Collin-Souffrin et al. (1986;  cf. Ferland et al. 1992 and
Shields et al. 1995) for details of the formation and radiation
transfer effects of various emission lines.}
Note that when we speak of an ``optically thick'' or
``optically thin'' cloud we are referring to the optical depth of the cloud to
hydrogen ionizing photons; the optical depth of a line, on the other hand,
refers to the optical depth of a certain cloud to the line.
The emerging pattern currently seems to be that, \emph{as \lratio\
increases, the EW of high-ionization or optically thick lines decreases
whereas the EW of low-ionization and optically thin lines increases. }

High-ionization lines are emitted from the illuminated surface of clouds;
optically thick lines come either from the illuminated surface (e.g., the
recombination line \lya) or from the thin transition layer of the partially
ionized \hi$^*$ region located immediately behind the hydrogen ionization
front.%
\footnote{For instance, \mgii\ $\lambda 2800$ originates only from
a thin layer with optical depth at the hydrogen Lyman limit
in the range 10 to 10$^4$ (Collin-Souffrin et al. 1986).}
By contrast, low-ionization, optically thin lines, such as the optical \feii\
multiplets, arise from the vast volume of the partially ionized \hi${^*}$ region
(i.e., from ionization-bounded clouds only).  Hence, the correlation patterns
described above may be telling us that, as \lratio\ increases, the hydrogen column
density (\colnh) of the clouds in the line-emitting region increases.  There
evidently exists some physical mechanism---closely linked with \lratio---that
regulates the global distribution of the properties of the clouds
gravitationally bound in the line-emitting region (including the inner NLR;
see Section 3.2).  Specifically, we propose (see also Dong et al. 2009a, their
Section 4) that \emph{there is a lower limit, set by \lratio, to the hydrogen column
density of the clouds gravitationally bound to the AGN line-emitting region}.
Low-\colnh\ clouds, even at small Eddington ratios (\lratio\ $\ll 1$), are
blown away because they are not massive enough to balance the radiation
pressure force, which is boosted by photoelectric absorption by at least an
order of magnitude (Fabian et al. 2006; Marconi et al. 2008).  According to
the photoionization calculations of Fabian et al. (2006, their Figure~1; see
also Ferland et al. 2009), which seem to be supported by observations (Fabian
et al. 2009), in dust-free clouds of $\colnh \gtrsim 10^{21}$ cm$^{-2}$ with
photoionization parameter $U \lesssim 1$ (valid for most AGNs),
the lower limit of the hydrogen column density of the clouds that
can survive in the BLR is approximately proportional to \lratio:
$\colnh > 10^{23} \, \lratio$~cm$^{-2}$.   The limit for dusty clouds is
higher, such that $\colnh > 5 \times 10^{23}$ \lratio\ cm$^{-2}$.  These
calculations suggest that AGNs with higher \lratio\ possess a larger
fraction of their line-emitting gas in high-\colnh\ clouds, conditions
that favor the production of low-ionization, optically thin lines such as \feii.

The above mechanism proposed to regulate \colnh\ explains the correlation
between \feii/\mgii\ and \lratio, but not the increase of EW(\feii) with
\lratio\ (see also Zhou et al. [2006] for a positive correlation between
EW(\feii) and $L_{5100}$),
since large-\colnh, \feii-emitting clouds are present whether
\lratio\ is high or low.  The positive correlation between  EW(\feii) and
\lratio\ requires that the absolute amount of line-emitting gas increases with
\lratio.  This is a natural expectation for any reasonable accretion scenario,
as \lratio\ scales with mass accretion rate.
On the other hand, the positive correlation between EW(\feii) and \lratio\
stands in sharp contrast with the behavior of \civ\ (Baskin \& Laor 2004) and
\mgii\ (Dong et al. 2009a), whose EWs \emph{decrease} with \lratio.  It is
not clear how these trends can be self-consistently explained in terms of
simple cloud physics.  In the case of \civ\ and \mgii, we can continue to
appeal to a change in the shape of the ionizing continuum with \lratio, one
of the more popular proposals to account for the classical Baldwin effect
(e.g., Zheng \& Malkan 1993; Korista et al. 1998).  However, this picture does
not offer any obvious solution for the opposite dependence of EW(\feii) on
\lratio.  This startling property of \feii\ strongly reinforces the
notion that it arises from regions that are physically distinct from the
bulk of the ``normal'' BLR, and that it is likely to be excited by
mechanisms other than photoionization.
In clouds of such high particle and column density as \feii\ emission favors,
some researchers have argued that photoionization heating might not
be sufficient to power the observed \feii\ line strengths.  An additional
source of heating, perhaps mechanical, may be necessary to enhance the
\hi$^{*}$ region (e.g., V{\'e}ron-Cetty et al. 2006; Joly et al. 2008).
Mechanical heating from outflows, whose strength increases with \lratio, might
be such a source, as there is marginal evidence that broad \mgii\ absorption
lines, presumably produced by outflows, are more frequently detected in QSOs
with stronger \feii\ emission (Zhang et al. 2010).
Since outflows are launched from the accretion disk and may have
large inclination angles, or may even be equatorial (Murray et al. 1995; Proga
et al. 2008), they have a high probability of colliding with clouds in the
line-emitting region and the torus.

In the picture proposed here to explain the observed emission-line correlations,
our focus has shifted from the detailed physics
(\emph{microphysics}) of the
accretion process of the central engine or of individual clouds, as was the
case in previous treatments (e.g., Netzer 1985; Zheng \& Malkan 1993; Korista
et al. 1998; Wandel 1999), to the statistical physics (\emph{macrophysics})
of the ensemble clouds instead (cf. Baldwin et al. 1995; Korista 1999; Dong et al. 2009b).

Lastly, we note that the strong, positive correlation between \feii/\mgii\ and
\lratio\ is unlikely to reflect any intrinsic relation between the Fe/Mg
abundance ratio and \lratio.  In most plausible scenarios that connect AGN
and starburst activity (e.g., Sanders et al. 1988; Davies et al. 2007), as
long as the delay between the two events is not more than 1 Gyr (the typical
timescale for chemical enrichment by Type Ia supernovae), we expect the
$\alpha$ elements to be enhanced relative to the iron-peak elements during the
active phase of the AGN.  We would thus expect the Fe/Mg abundance ratio to
correlate {\em negatively} with \lratio, which is opposite to the trend seen.

\subsection{The Sites of the \feii--emitting Regions}
As reported in the companion paper by Dong et al. (2010), narrow \feii\ emission
is prevalent in type~1 AGNs, yet not present at all in type~2 AGNs.  We suggest
that narrow \feii\ emission arises from gas in the innermost regions
of the NLR located interior to the obscuring torus, in the so-called inner NLR
or intermediate-line region proposed previously by some researchers
(see references in Section 2.1.2).
This is further supported by the strong correlation between the strength of
narrow \feii\ and \lratio\ found in the present study, which suggests that the
region emitting narrow \feii\ is probably rather homogeneous and well-defined.
It has been estimated that the torus has an inner edge of a few parsecs,
roughly the expected location of the dust sublimation radius
(Barvainis 1987; Suganuma et al. 2006, and references therein), and a total
radial extent of several tens of parsecs
(e.g., Kl{\"o}ckner et al. 2003; Jaffe et al. 2004;
see Granato \& Danese 1994 for a model).
As a low-ionization species, \feii\ may preferentially avoid
the ionization cone and be largely confined to a disk-like geometry
along the plane of the torus. Because the optical \feii\ lines are
emitted most efficiently at high densities ($\sim 10^6-10^8$ cm$^{-3}$;
Verner et al. 2000;
V\'eron-Cetty et al. 2004), it is likely to be concentrated toward the inner
NLR.  This is unlike the case of high-ionization, high-critical density narrow
emission lines such as \oiii\,$\lambda5007$, which has a significant component
in the inner NLR but lies preferentially in the ionization cone
(Schmitt et al. 2003; Zhang et al. 2008, and references therein).
The lack of correlation between narrow \feii\ and \oiii\ (Section 2.2)
supports the notion that they arise from distinct emission regions.

In contrast to the narrow-line emission, the EW of the optical broad \feii\
emission shows only a moderate correlation with \lratio.  What causes this
different behavior?  A possible explanation is that broad \feii\ does not
originate from a unique site---that is, not only from the clouds bound to the
BLR---but from a mixture of different locations.
This interpretation is consistent with the reverberation mapping observations
of \feii, whose broad and flat-topped cross-correlation function (Vestergaard
\& Peterson 2005; Kuehn et al. 2008) indicates an extended emitting region.
A plausible additional site for the formation of broad \feii\ emission and
other low-ionization lines (e.g., Balmer lines and \mgii) is the surface of
the accretion disk (Collin-Souffrin et al. 1980; Collin-Souffrin 1987; Murray
\& Chiang 1997; Zhang et al. 2006), on scales of a few hundred
gravitational radii ($\rg \equiv GM/c^2$).  This is
convincingly supported by the detection of low-ionization lines with
double-peaked profiles in a minority of AGNs (Balmer lines: Chen et al. 1989,
Eracleous \& Halpern 2003, Strateva et al. 2003; \mgii: Halpern et al. 1996),
particularly by the discovery of Balmer lines having, in addition to a
symmetric broad component located at the system velocity, a separate,
extremely broad double-peaked component that is apparently gravitationally
redshifted (Chen et al. 1989; Wang et al. 2005; see Wu et al. 2008 for a
treatment of the general case of a twisted, warped disk).  This disk component
contributes partly to the often-called ``red shelf'' or ``very broad
component'' (see, e.g., Sulentic et al. 2000a) of the Balmer line profile.
While double-peaked profiles are neither a necessary nor a sufficient condition
for a disk origin, we note that double-peaked \feii\ emission lines are
observed in other disk-accreting systems such as cataclysmic variable stars,
which, according to Doppler tomography, definitely arise from an
accretion disk (e.g., Roelofs et al. 2006; Smith et al. 2006).  Note that the
accretion disk radii that can emit low-ionization lines are smaller
than the critical radius of the gravitationally unstable part of the disk,
$r_{\rm crit}\approx 2 \times 10^4 \, (\mbh/10^7\msun)^{-0.46}\,\rg$
(Collin \& Hur{\'e} 2001).  Collin \& Hur{\'e} (2001; see also Leighly 2004)
suggest that \emph{the gravitationally unstable region of the disk is a source
of clouds for the line-emitting region of AGNs, particularly for
low-ionization species such as \mgii\ and {Fe\,{\footnotesize II}} discussed
in this paper.}
Other candidate sites for broad \feii\ emission might be outflows, which are
prone to fragmentation (Proga et al. 2008), and gas infalling from the torus
(Gaskell \& Goosmann 2008; Hu et al. 2008a, 2008b).
Compared to the gas in the torus, these clouds are located on scales smaller than
the dust sublimation radius and are believed to have little or no dust content
(see Suganuma et al. 2006; Elitzur \& Shlosman 2006, and references therein).


\section{Conclusions and Implications}

We find compelling evidence that narrow \feii\ emission originates from a
well-defined location, which we speculate to be the inner NLR on scales smaller
than the torus.  The lack of correlation between the strengths of narrow
\feii\ and \oiii\,$\lambda5007$ suggests that they are emitted from different
regions.  On the other hand, consistent with the findings from reverberation
mapping studies (Vestergaard \& Peterson 2005; Kuehn et al. 2008), the sites
of broad \feii\ emission are likely to be more diverse.  We speculate that,
similar to the situation for broad \hb, a significant fraction of the broad
\feii\ emission may originate from the surface of the accretion disk instead
of from clouds gravitationally bound to the BLR.  It appears that \feii\ emission can arise
from any gas surrounding the central engine of AGNs that has sufficiently high
particle density, column density, and heating energy input.

Although the excitation mechanisms of \feii\ emission are complex, and the
sites of line formation still poorly known,
the relative strength of (optical) \feii\ emission
with respect to both the continuum and other emission lines (particularly \mgii)
correlate strongly and positively with \lratio, and likely \emph{not}
with AGN luminosity or \mbh\ intrinsically.  Their
apparent correlations with luminosity and \mbh\ are a secondary
effect of their much stronger correlations with \lratio.%
\footnote{UV \feii\ emission is a perplexing exception; its
EW does not correlate with \lratio.
The formation and radiative transfer of the UV \feii\ lines are poorly
understood.  We refer readers to several recent papers by Bruhweiler \& Verner
(2008), Ferland et al. (2009), and Sameshima et al. (2011).}
Combined with previous findings on
\civ\,$\lambda1549$ and \mgii\,$\lambda2800$ (Bachev et al. 2004; Baskin \&
Laor 2004; Dong et al. 2009a; cf. Shemmer et al. 2004; Warner et al. 2004;
Dong et al. 2009b), this means that, apart from the zeroth-order global
similarity in QSO spectra, the first-order variation of 
these important emission lines,
either high-ionization or low-ionization, optically thick or optically thin,
is controlled by \lratio. We attribute the essential underlying
physical mechanism of these correlations to the overall gas supply increasing
with \lratio, as well as the role \lratio\ plays in regulating the
distribution of hydrogen column density of the clouds gravitationally
bound to the line-emitting regions.
Specifically, we can conclude that the Baldwin effect of \civ, \mgii\ and
optical \feii\ is driven by \lratio\ (see also Bachev et al. 2004; Baskin \&
Laor 2004; Dong et al. 2009a).

If the observed large scatter of \feii/\mgii\ at any given redshift
predominantly reflects the spread in \lratio\ of the QSO population, then
there is still hope that \feii/\mgii\ can be used as a measure of the Fe/Mg
abundance ratio to study chemical evolution in AGN environments once the
systematic variation caused by \lratio\ is corrected according to the
empirical relations (Eqn. 1--3) presented in this paper.  In this respect,
since their relation with \lratio\ is relatively tighter, the optical \feii\
features, particularly the narrow component, might be more effective than the
usually used UV \feii\ features.

\acknowledgments
We thank the anonymous referee for his/her thorough, critical, and helpful
comments.  X.-B.\,D. thanks Kirk Korista, Martin Gaskell, Aaron Barth,
Alessandro Marconi, Philippe V\'{e}ron, Monique Joly,
Daniel Proga, Xueguang Zhang, and Feng Yuan for helpful discussions and comments.
He is grateful to Sheng-Miao Wu, Lei Chen, and Fu-Guo Xie for their warm hospitality and
brainstorming discussions during visits to Shanghai Observatory.  We thank
Zhen-Ya Zheng for help in improving the IDL figures,
and Yifei Chen, Qian Long and Paul Collison for computing support.
This work is supported by Chinese NSF grants NSF-10703006,
NSF-10728307, NSF-10973013, NSF-11033007, and NSF-11073019, a CAS
Knowledge Innovation Program (Grant No. KJCX2-YW-T05),
a National 973 Project of China (2009CB824800), 
and the Fundamental Research Funds for the Central Universities
(USTC WK2030220004).
The research of L.C.H. is supported by the
Carnegie Institution for Science.  The visit of X.-B.\,D. to Carnegie
Observatories is mainly supported by OATF, USTC (up to April 2010).
Funding for the SDSS and SDSS-II has been provided by the Alfred P.
Sloan Foundation, the Participating Institutions, the National
Science Foundation, the U.S. Department of Energy, the National
Aeronautics and Space Administration, the Japanese Monbukagakusho,
the Max Planck Society, and the Higher Education Funding Council for
England.  The SDSS Web Site is http://www.sdss.org/. The SDSS is
managed by the Astrophysical Research Consortium for the
Participating Institutions. The Participating Institutions are the
American Museum of Natural History, Astrophysical Institute Potsdam,
University of Basel, University of Cambridge, Case Western Reserve
University, University of Chicago, Drexel University, Fermilab, the
Institute for Advanced Study, the Japan Participation Group, Johns
Hopkins University, the Joint Institute for Nuclear Astrophysics,
the Kavli Institute for Particle Astrophysics and Cosmology, the
Korean Scientist Group, the Chinese Academy of Sciences (LAMOST),
Los Alamos National Laboratory, the Max-Planck-Institute for
Astronomy (MPIA), the Max-Planck-Institute for Astrophysics (MPA),
New Mexico State University, Ohio State University, University of
Pittsburgh, University of Portsmouth, Princeton University, the
United States Naval Observatory, and the University of Washington.

\appendix
\section{Selection Criterion to Minimize Contamination from Host Galaxy Starlight}
The SDSS spectra were taken through a 3\arcsec-diameter fiber aperture, which
corresponds to $\sim 17.4$ kpc at the mean redshift of the full sample
($z=0.46$), and thus generally includes a large amount of host galaxy
starlight.  To ensure that starlight does not dominate the total flux and
significantly impact the measurement of the AGN continuum and the EWs of
emission lines, we design a selection criterion that requires that the EWs of
the \caii\,K (3934 \AA), \caii\,H~+~H$\epsilon$ (3970 \AA), and H$\delta$
(4102 \AA) absorption features be undetected at $2\,\sigma$ significance.
We calculate the EWs of these absorption features following Balogh et al.
(1999), by summing the observed flux in a narrow wavelength range centered on
each line above/below the local continuum level.  The continuum is determined
by linearly fitting line-free windows placed on either side of the line.
As our goal is not to measure accurately the absorption-line EWs but instead
to detect and eliminate objects with significant absorption lines, we
place the continuum windows as close as possible to the absorption features,
choosing a relatively narrow wavelength range to minimize the contamination
from nearby emission lines.  The [Fe\,V] $\lambda 4072$ emission line is
masked out from the continuum window blueward of H$\delta$.  The wavelength
windows for the lines and local continua are listed in Table A1.
The measurement uncertainties of the EWs are estimated in a similar way as
Balogh et al. (1999), accounting for both the error in the continuum fit and
the error in every pixel comprising the line.  Generally, the \caii\,H~+~H$\epsilon$
or H$\delta$ features are only auxiliary, and \caii\,K alone can
effectively detect starlight because this feature is free from contamination
by emission lines.  Nevertheless, in AGN-dominated spectra the measurement of
\caii\,K absorption can be inaccurate because of emission-line contamination
in its continuum windows.  Thus, for objects that have \caii\,K absorption
detected at $\geq 2\,\sigma$ significance but \caii\,H~+~H$\epsilon$ or H$\delta$
detected at $< 1\,\sigma$, we inspect them visually to avoid false
detection of \caii\,K absorption.
Roughly 10\% of the 4178 sources in our sample
are picked up this way; such a small fraction do not impact seriously
the efficiency of the above selection criterion.

We performed simulations to test the effect of our selection criterion on
limiting starlight contamination.  We select SDSS spectra of 50 QSOs at
$z<1.16$ with monochromatic luminosity at 4200 \AA\ $> 5 \times 10^{45}$ \lum,
and 200 spectra of galaxies with little or no emission lines.  We build
artificial spectra by adding together galaxy and QSO spectra
with various starlight contributions.  Applying our absorption-line
selection to the artificial spectra, we find that
the detected fractions of QSOs are 99\%, 83\%, 21\%, and 8\% for
spectra with starlight contributions at 4200~\AA\ of
5\%, 10\%, 15\%, and 20\%, respectively.
Thus, our selection criterion corresponds to a starlight contribution of
$\lesssim 10$\% around 4200~\AA, which introduces, at most,
0.002~dex (0.5\%) to the variance (namely, the square of the 1-$\sigma$ error)
of the emission-line EWs.

\begin{deluxetable}{l|ccc}
\centering
\tablenum{A1: Definition of Stellar Absorption-line Equivalent Widths}
\tablehead{
\colhead{Index}  & \colhead{Blue Continuum}  & \colhead{Line}  & \colhead{Red Continuum}
}
\startdata
\caii\,K                 &  3899--3919 &    3919--3952 & 3990--4010  \\
\caii\,H~+~H$\epsilon$   &  3899--3919 &    3957--3985 & 3990--4010  \\
H$\delta$             &  4046--4066,~4076--4086 &    4086--4120 & 4120--4150
\enddata
\tablecomments{Wavelengths are measured in vacuum, in units of \AA.}
\end{deluxetable}


\clearpage

\setcounter{figure}{0}
\setcounter{table}{0}

\figurenum{1}
\begin{figure}[tbp]
\epsscale{1}
\label{fig:optfitting}
\plotone{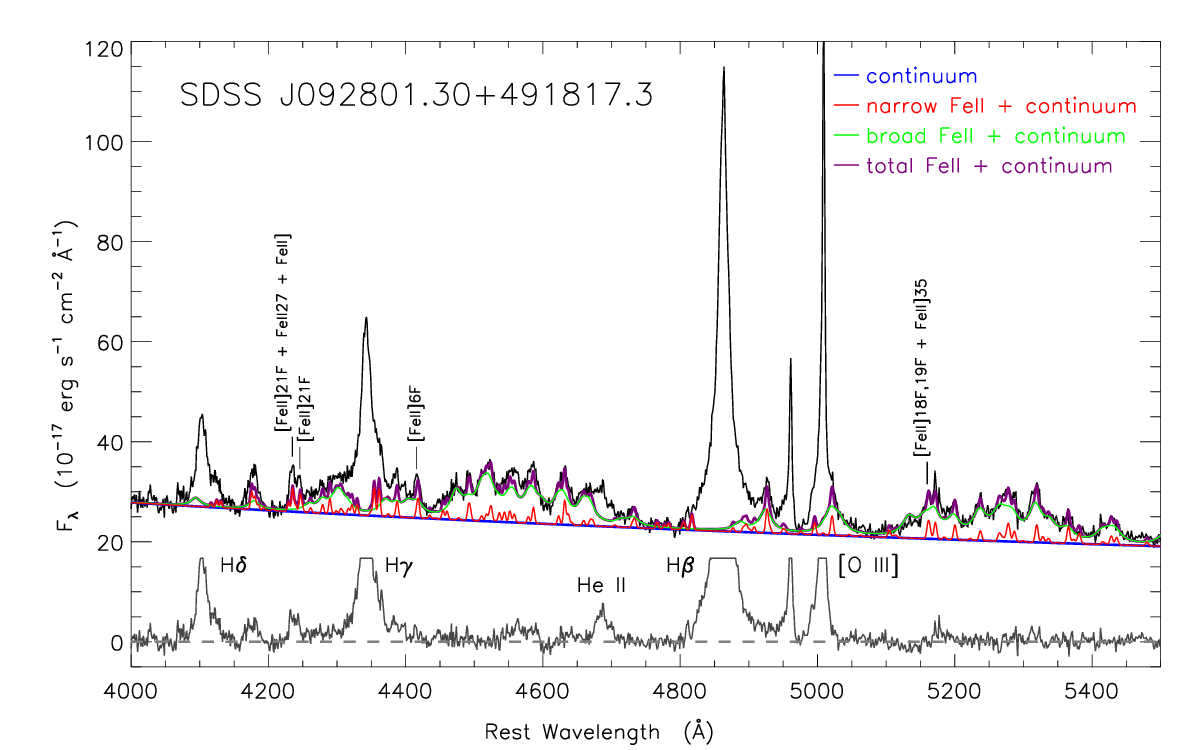}\caption{Demonstration of
spectral fitting in the optical region. We show
the SDSS spectrum (black),
the local AGN continuum (blue),
the continuum plus narrow-line \feii\ emission (red),
the continuum plus broad-line \feii\ emission (green),
the pseudocontinuum (continuum plus total \feii, purple),
and the pseudocontinuum-subtracted residuals (gray).
Also marked are some narrow \feii\ lines
(particularly the forbidden lines) that are sharp and distinct from nearby
broad \feii\ lines.
The strong emission lines in the residual spectrum are truncated for clarity.
}
\end{figure}

\figurenum{2}
\begin{figure}[tbp]
\epsscale{0.9}
\label{fig:demoobjs}
\plotone{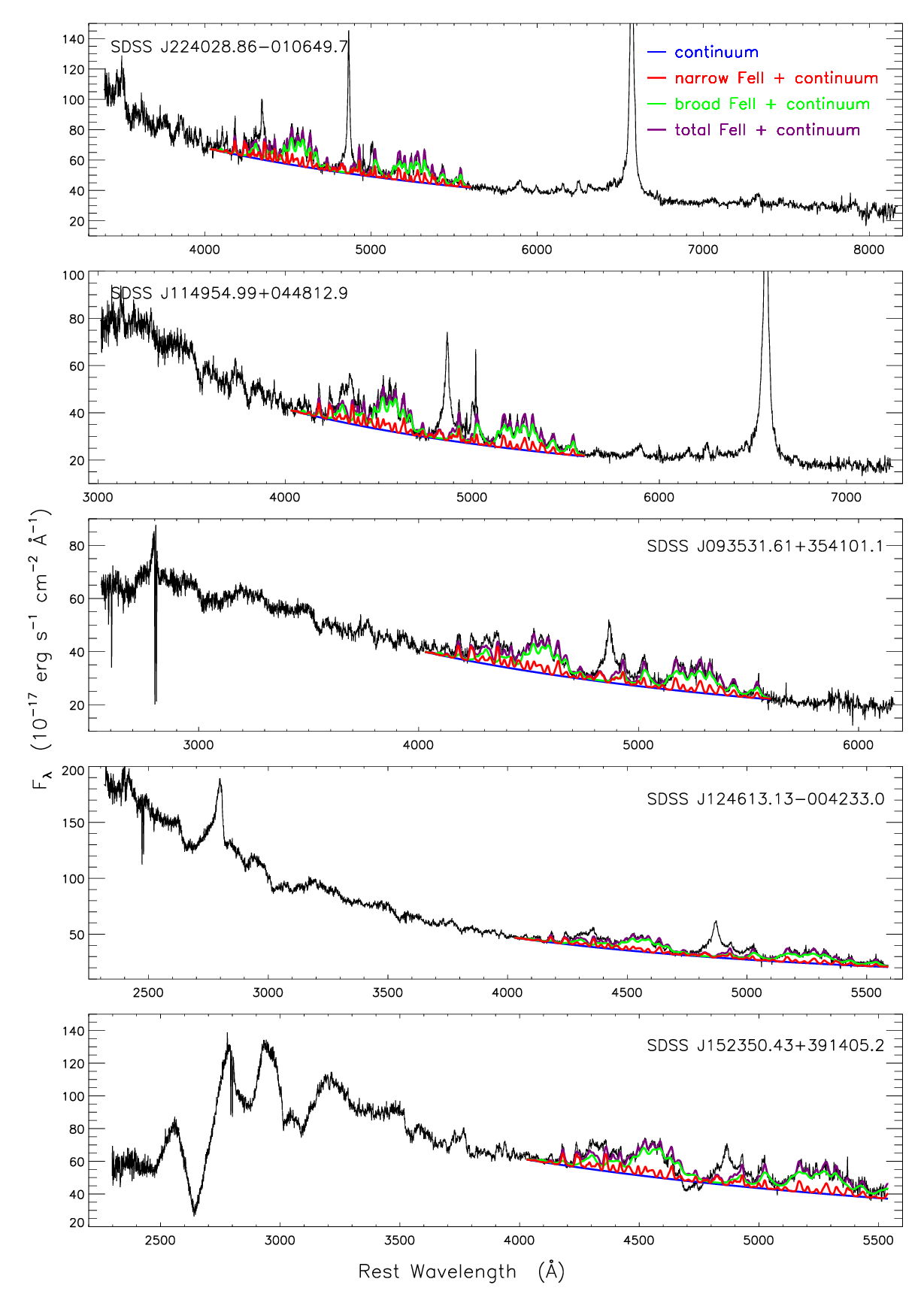}
\caption{Examples of SDSS spectra of different classes of AGNs with
strong narrow \feii\ emission.
Also plotted is our continuum fitting in the rest-frame wavelength range
4000--5600 \AA. Individual spectral components are denoted as in Figure 1.
}
\end{figure}

\figurenum{3}
\begin{figure}[tbp]
\epsscale{1}
\label{fig:uvfitting}
\plotone{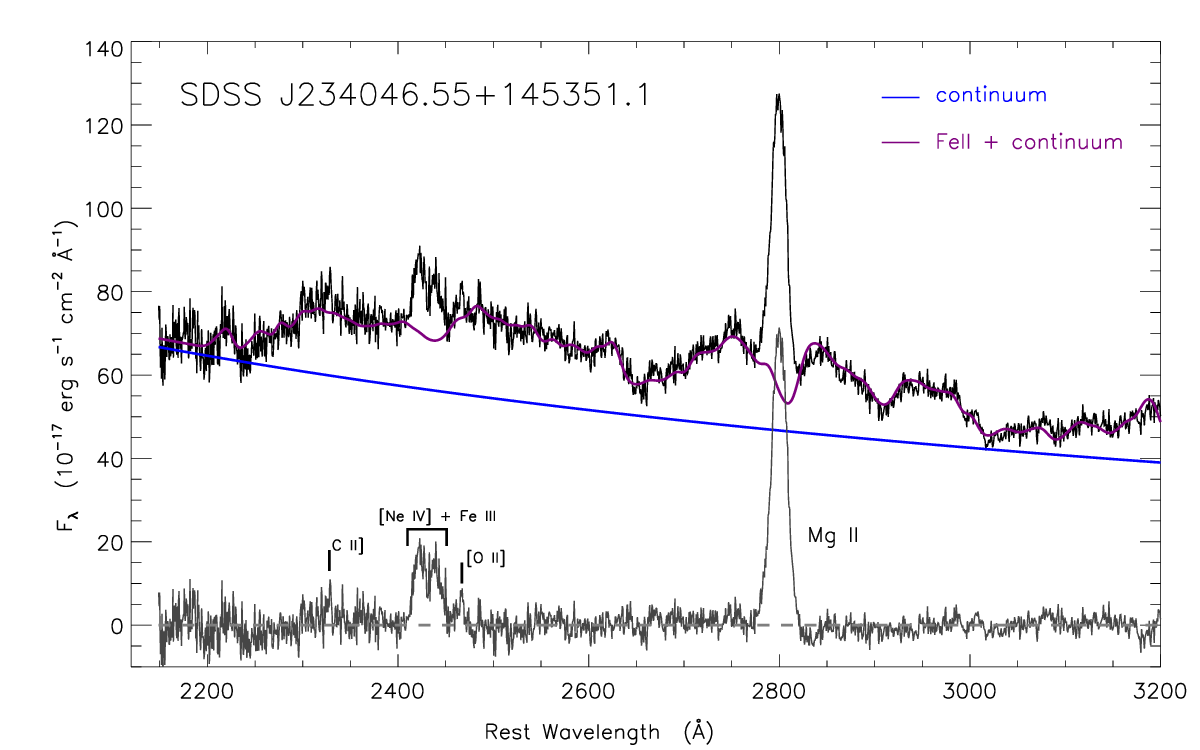}\caption{Demonstration of the continuum
fitting in the UV region. We show the SDSS spectrum (black),
the local AGN continuum (blue),
the pseudocontinuum (continuum plus \feii\ emission, purple),
and the pseudocontinuum-subtracted residuals (gray).
}
\end{figure}

\figurenum{4}
\begin{figure}[tbp]
\epsscale{1} \label{fig:testschemes} \plotone{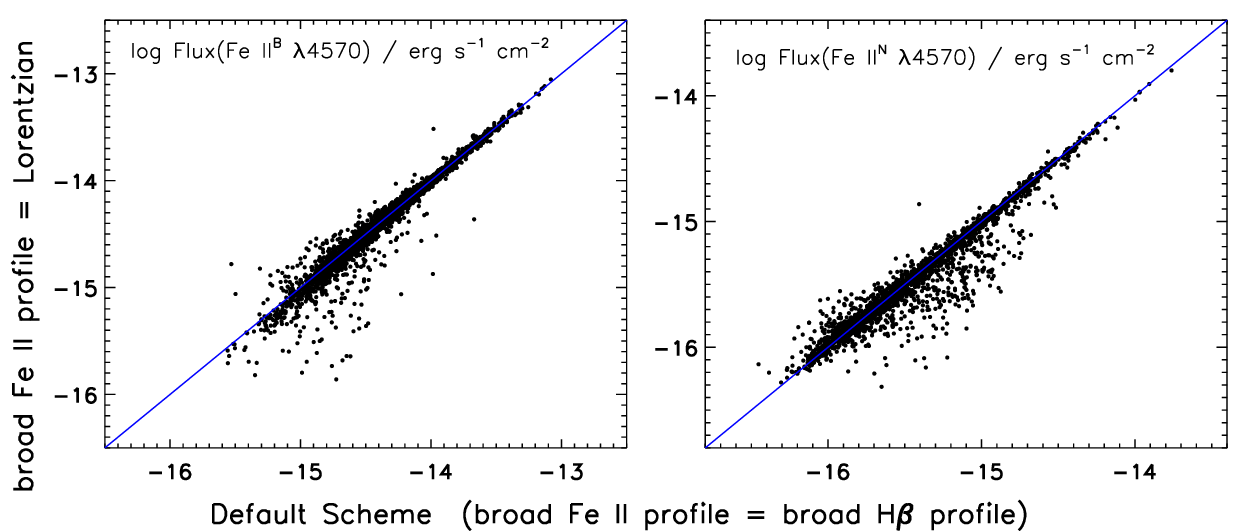} %
\caption{The fluxes of broad (left) and narrow (right)
\feii\,$\lambda4570$ emission detected at $>3 \,\sigma$
significance, calculated from the best fits using two different
schemes: the default scheme where the profile of broad \feii\ is set
to that of broad \hb, and another where broad \feii\ is modeled as a
Lorentzian with variable width. The $1\,\sigma$ relative errors for
the fluxes of broad and narrow \feii\,$\lambda4570$ are only 11\%
and 17\%, respectively.  These errors are estimated from the bootstrap
method and do not account for the uncertainties caused by possible
\feii\ template mismatch.}
\end{figure}

\figurenum{5}
\begin{figure}[tbp]
\epsscale{1}
\label{fig:ewb4570vsuvfeii}
\plotone{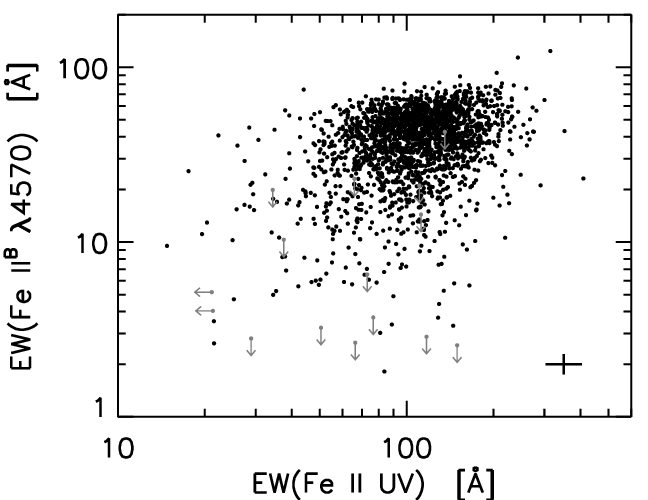}
\caption{Distribution of EWs of broad \feii\,$\lambda4570$
and UV \feii\ emission.
Black dots denote reliable detections at $\geq 2\sigma$ significance;
gray arrows give upper limits (see Section 2.1.2).
The bottom-right corner shows a representative error bar,
the length of which corresponds to the 1\,$\sigma$ total error, which
includes contributions from uncertainties arising from line deblending
and \feii\ fitting (see Section 2.1.3).
}
\end{figure}

\figurenum{6}
\begin{figure}[tbp]
\epsscale{1}
\label{fig:fluxb4570vsuvfeii}
\plottwo{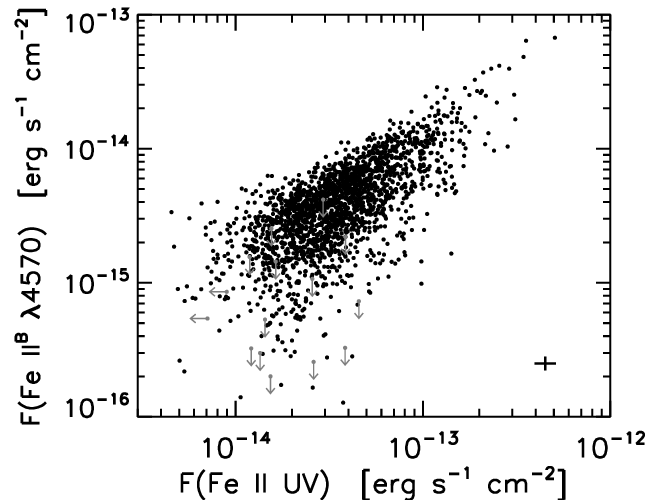}{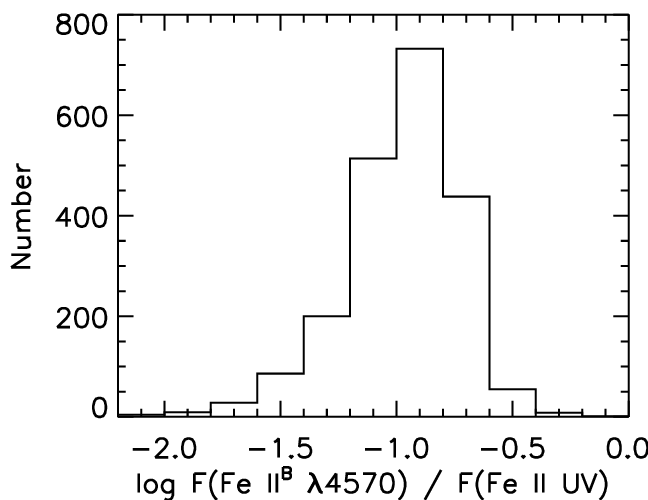}
\caption{Left: Distribution of fluxes of broad \feii\,$\lambda4570$
and UV \feii\ emission. The symbols are the same as in Figure~5.
Right: Histogram of the flux ratios of broad \feii\,$\lambda4570$ to UV \feii\ emission,
for the objects with both emission lines detected at $\geq 2\sigma$ significance.
}
\end{figure}

\figurenum{7}
\begin{figure}[tbp]
\epsscale{1}
\label{fig:uvfeiivsbmgii}
\plottwo{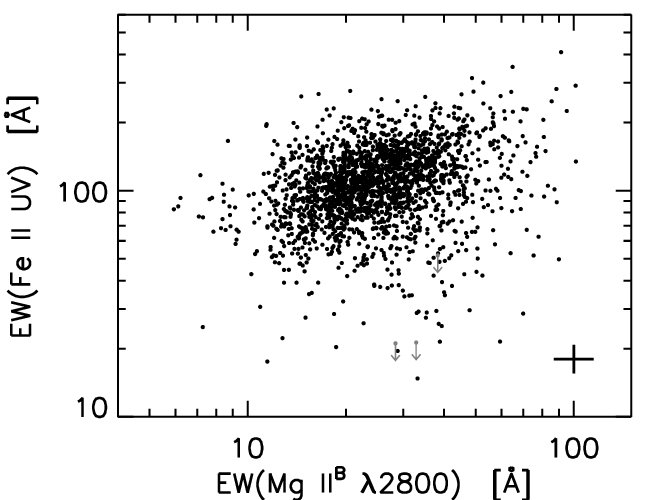}{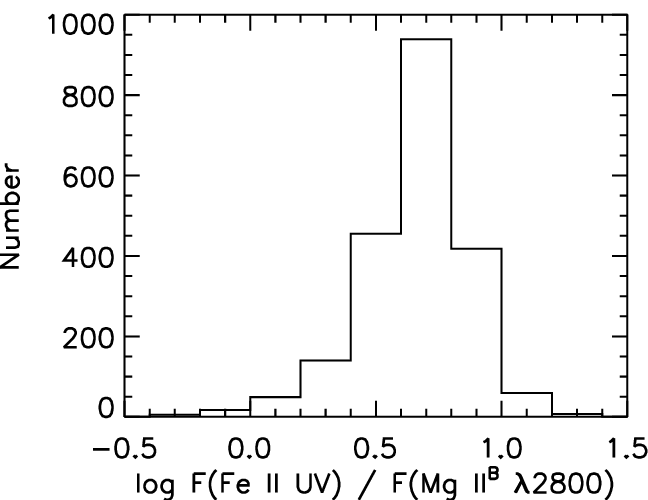}
\caption{Left: Distribution of EWs of UV \feii\ and
broad \mgii\,$\lambda2800$. The symbols are the same as in Figure~5.
Right: Histogram of the flux ratios of UV \feii\ to broad \mgii\,$\lambda2800$,
for the objects with both emission lines detected at $\geq 2\sigma$ significance.
}
\end{figure}

\figurenum{8}
\begin{figure}[tbp]
\epsscale{1}
\label{fig:b4570vsbhb}
\plottwo{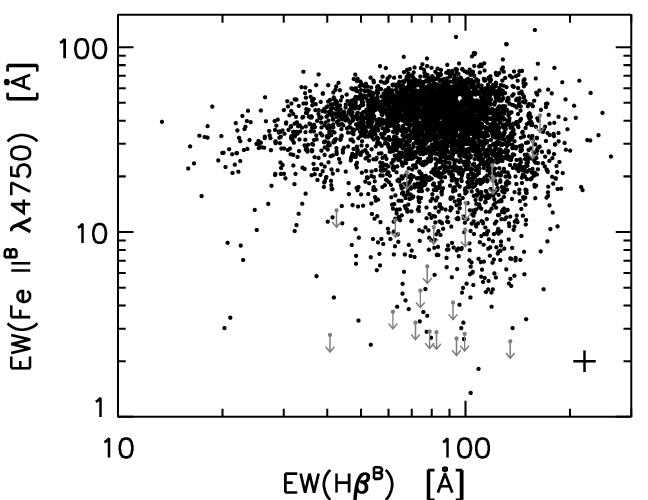}{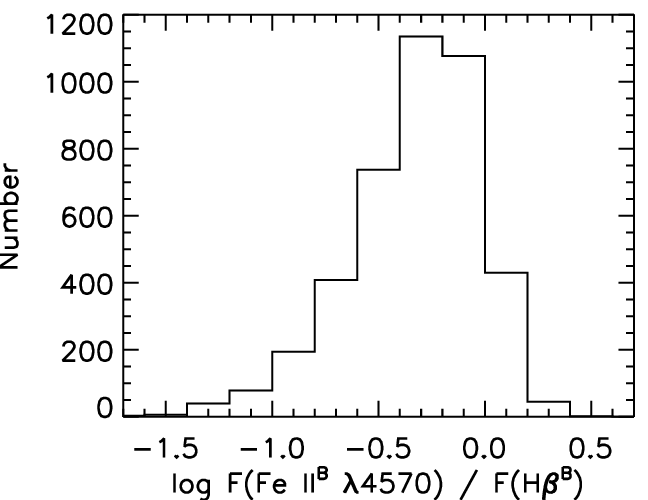}
\caption{Left: Distribution of EWs of broad \feii\,$\lambda4570$ and
broad \hb. The symbols are the same as in Figure~5.
Right: Histogram of the flux ratios of broad \feii\,$\lambda4570$ to broad \hb,
for the objects with both emission lines detected at $\geq 2\sigma$ significance.
}
\end{figure}

\figurenum{9}
\begin{figure}[tbp]
\epsscale{0.6}
\label{fig:n4570vsoiii}
\plotone{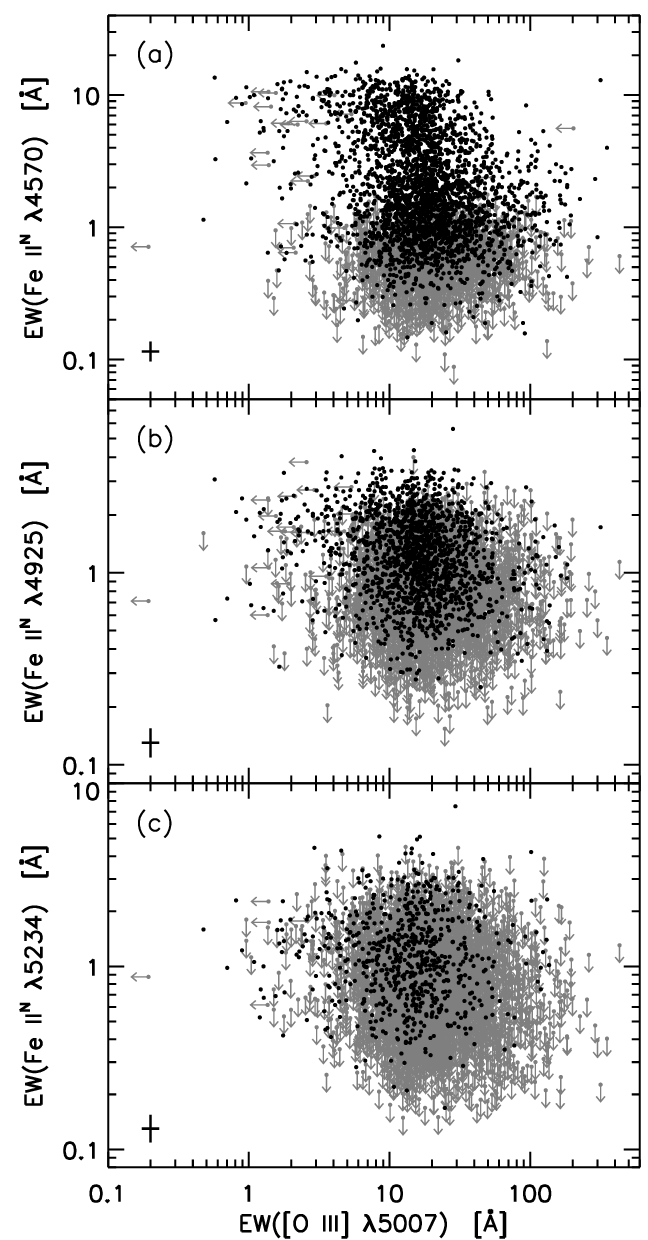}
\caption{Distribution of EWs of narrow (a) \feii\,$\lambda4570$,
(b)  \feii\,$\lambda4925$, and (c)  \feii\,$\lambda5234$ versus that of
\oiii\,$\lambda5007$. \feii\,$\lambda4570$ and \oiii\,$\lambda5007$ are
measured from the best-fit model, while \feii\,$\lambda4925$ and \feii\,$\lambda5234$ are measured from the residual
spectra after the continuum, broad \feii, and other broad emission lines
are subtracted. The symbols are the same as in Figure~5.
}
\end{figure}

\figurenum{10}
\begin{figure}[tbp]
\epsscale{1}
\label{fig:blewcorrs}
\plotone{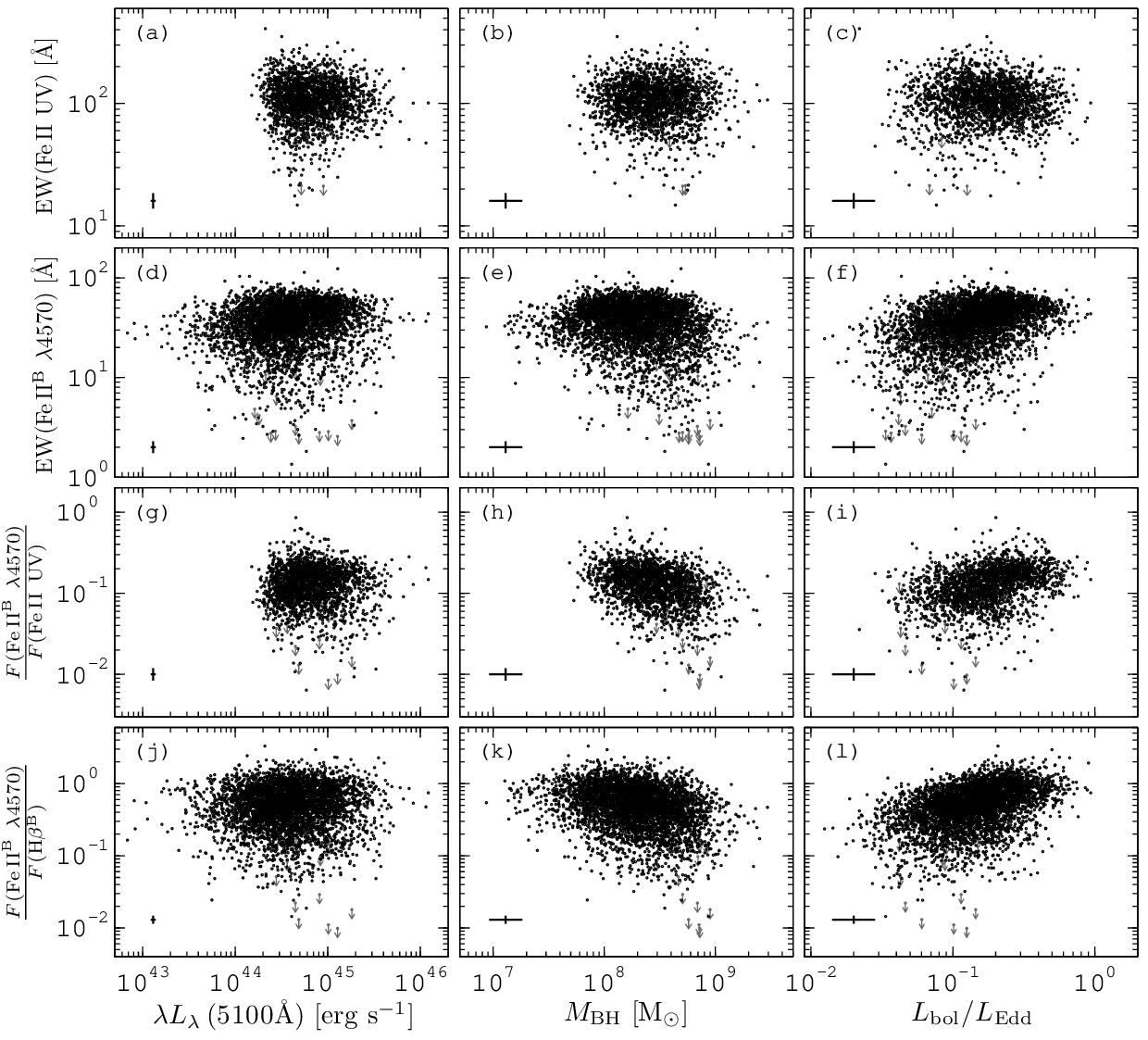}
\caption{
The dependence of broad \feii\ emission on
continuum luminosity $\lambda L_{\lambda}$(5100 \AA), BH mass (\mbh),
and Eddington ratio ($\lbol/\ledd$). The symbols are the same as in Figure~5.}
\end{figure}
\clearpage

\figurenum{11}
\begin{figure}[tbp]
\epsscale{1}
\label{fig:nlewcorrs}
\plotone{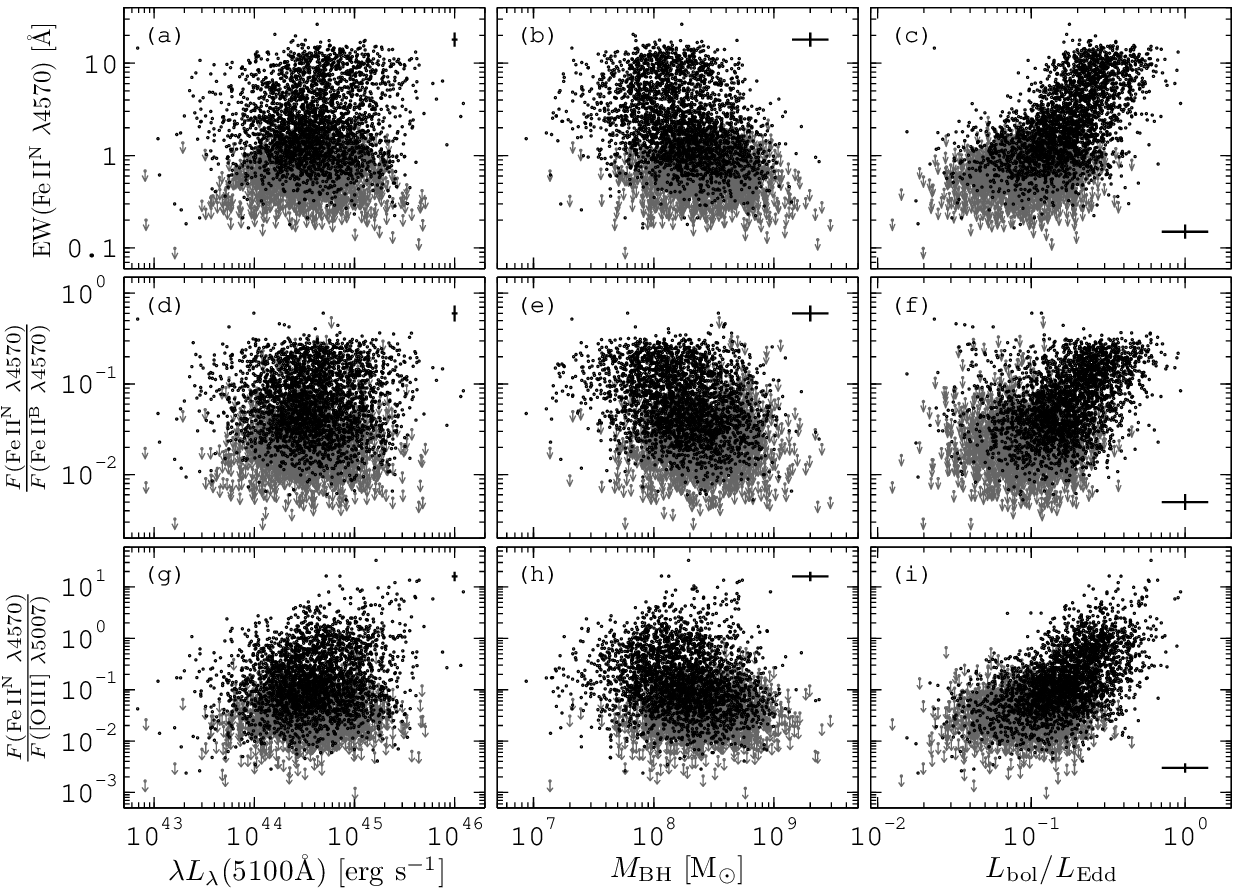}
\caption{
The dependence of narrow \feii\ emission on
continuum luminosity $\lambda L_{\lambda}$(5100 \AA), BH mass (\mbh),
and Eddington ratio ($\lbol/\ledd$). The symbols are the same as in Figure~5.}
\end{figure}

\figurenum{12}
\begin{figure}[tbp]
\epsscale{0.6}
\label{fig:femgell} \plotone{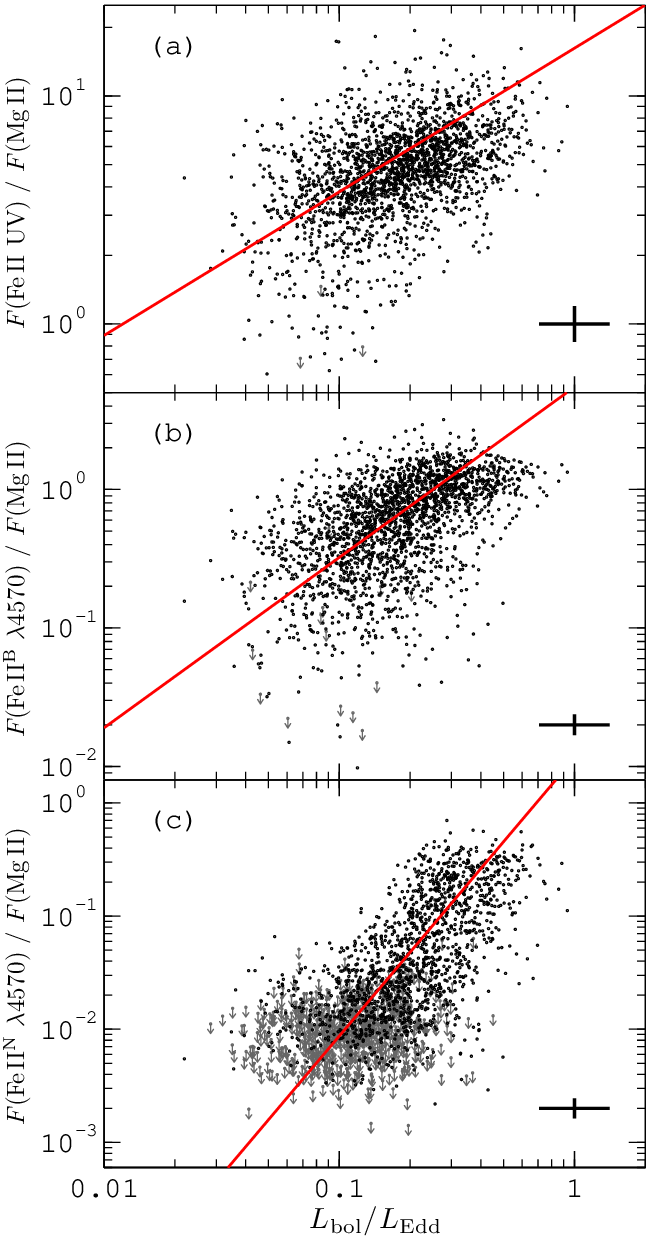}
\caption{ The strength of broad (UV and optical) and narrow \feii\
emission relative to broad \mgii\,$\lambda2800$ as a function of
Eddington ratio.  Also plotted are the best-fitting linear relations
in log--log scale (see Eqn. 1--3). The symbols are the same as in Figure~5.}
\end{figure}

\clearpage

\begin{sidewaystable*}

\topmargin 0.0cm \evensidemargin = 0mm \oddsidemargin = 0mm
\tiny
\caption{Optical Continuum and Emission-line Parameters of the Full Sample}
\label{table1}

\begin{tabular}{lccccccccccccc}
 \hline \hline
SDSS Name                    & $z$     &
$\log L_{5100}$              & $\beta$ &
FWHM(H$\beta^{\rm B}$)       &
$\log F$(Fe\,II$^{\rm N}$)   &
EW(Fe\,II$^{\rm N}$)         &
$\log F$(Fe\,II$^{\rm B}$)   &
EW(Fe\,II$^{\rm B}$)         &
$\log F$(H$\beta^{\rm N}$)   &
$\log F$(H$\beta^{\rm B}$)   &
EW(H$\beta^{\rm B}$)         &
$\log F$([O\,III]\,$\lambda5007$)       &
EW([O\,III]\,$\lambda5007$)             \\
(1)  & (2)  & (3)  & (4)  & (5)  &
(6)  & (7)  & (8)  & (9)  & (10) &
(11) & (12) & (13)  & (14)  \\
\hline
J000011.96$+$000225.2  &  0.4790  &  44.69  &  $-$2.45  &  3034    &   $ -$14.96   &   $ $7.6     &    $-$13.99   &  71.5   &      $-$15.68    &  $-$13.89   &   104.9  &  $ -$14.87    &    $ $11.7      \\     
J000043.95$-$091134.9  &  0.4388  &  44.62  &  $-$1.59  &  5973    &   $<-$16.16   &   $<$0.5     &    $-$14.34   &  33.4   &      $-$16.00    &  $-$14.29   &   41.1   &  $ -$15.02    &    $ $8.1       \\     
J000102.19$-$102326.8  &  0.2943  &  44.20  &  $-$1.00  &  7748    &   $ -$16.04   &   $ $0.7     &    $-$14.98   &  8.2    &      $-$15.16    &  $-$14.02   &   81.3   &  $ -$14.20    &    $ $55.4      \\     
J000109.14$-$004121.5  &  0.4166  &  44.32  &  $-$1.62  &  1899    &   $ -$16.09   &   $ $1.0     &    $-$14.67   &  27.2   &      $-$15.52    &  $-$14.60   &   35.1   &  $ -$14.99    &    $ $15.0      \\     
J000110.96$-$105247.4  &  0.5283  &  44.98  &  $-$2.23  &  6806    &   $<-$16.13   &   $<$0.3     &    $-$14.11   &  35.3   &      $-$15.53    &  $-$13.73   &   97.2   &  $ -$14.52    &    $ $16.8      \\     
J000111.21$-$002011.2  &  0.5178  &  44.51  &  $-$1.63  &  3465    &   $ -$15.55   &   $ $3.8     &    $-$14.47   &  46.3   &      $-$15.80    &  $-$14.14   &   110.1  &  $ -$14.54    &    $ $46.2      \\     
J000115.99$+$141123.0  &  0.4037  &  44.38  &  $-$1.37  &  5178    &   $<-$16.15   &   $<$0.7     &    $-$14.98   &  11.2   &      $-$16.42    &  $-$14.13   &   85.6   &  $ -$14.36    &    $ $53.0      \\     
\hline
\end{tabular}
\medskip
\vfill
{\normalsize Note. ---
Column (1) official SDSS name;
Column (2) redshift measured by the SDSS pipeline;
Column (3) luminosity of the power-law continuum at 5100\,\AA, $\lambda L_{\lambda}$(5100 \AA);
column (4) local continuum slope fitted in the rest-frame wavelength range of 4000--5600 \AA\
($f_{\lambda} = \lambda ^{\beta}$);
Column (5) FWHM of broad H$\beta$, corrected for instrumental broadening;
Column (6) flux of narrow Fe\,II $\lambda4570$
(integrated in the range of 4434--4684 \AA\ from the best-fit model);
Column (7) rest-frame EW of narrow Fe\,II $\lambda4570$;
Column (8) flux of broad Fe\,II $\lambda4570$
(integrated in the range of 4434--4684 \AA\ from the best-fit model);
Column (9) rest-frame EW of broad Fe\,II $\lambda4570$;
Column (10) flux of the narrow component of H$\beta$;
Column (11) flux of the broad component of H$\beta$;
Column (12) rest-frame EW of the broad component of H$\beta$;
Column (13) flux of [O\,III] $\lambda5007$;
Column (14) rest-frame EW of [O\,III] $\lambda5007$.
Luminosities, fluxes, EWs, and FWHM are in units of \lum, \flux, \AA, and \kms,
respectively. (This table is available in its entirety in a machine-readable form in the online
journal. A portion is shown here for guidance regarding its form and content.)}

\end{sidewaystable*}
\clearpage

\begin{deluxetable}{lccccccc}
\tabletypesize{\tiny}
\tablecaption{Near-UV Continuum and Emission-line Parameters of the UV Subsample}
\tablehead{
\colhead{SDSS Name} &
\colhead{$\log L_{3000}$} &
\colhead{FWHM(Mg\,II$^{\rm B}$)} &
\colhead{$\log F$(UV Fe\,II)} &
\colhead{EW(UV Fe\,II)} &
\colhead{$\log F$(Mg\,II$^{\rm N}$)} &
\colhead{$\log F$(Mg\,II$^{\rm B}$)} &
\colhead{EW(Mg\,II$^{\rm B}$)} \\
\colhead{(1)}  & \colhead{(2)} & \colhead{(3)} & \colhead{(4)} & \colhead{(5)} &
\colhead{(6)}  & \colhead{(7)} & \colhead{(8)}
}
\startdata
J000011.96$+$000225.2  & 45.02  &   2898   &   $ -$13.04   &   $ $180.6    &   $-$15.81   &  $-$13.79   &  35.4   \\
J000110.96$-$105247.4  & 45.24  &   6135   &   $ -$13.33   &   $ $77.3     &   $-$15.06   &  $-$13.90   &  22.4   \\
J000111.21$-$002011.2  & 44.71  &   2601   &   $ -$13.77   &   $ $93.3     &   $-$15.44   &  $-$14.46   &  20.1   \\
J000559.20$+$153125.1  & 44.65  &   3014   &   $ -$13.38   &   $ $179.5    &   $-$15.88   &  $-$13.98   &  50.2   \\
J000945.46$+$001337.1  & 45.12  &   2342   &   $ -$13.46   &   $ $149.3    &   $-$15.99   &  $-$14.32   &  22.0   \\
J001024.22$+$153331.3  & 45.69  &   3317   &   $ -$13.08   &   $ $102.0    &   $-$15.72   &  $-$13.92   &  15.9   \\
J001104.84$-$092357.8  & 45.08  &   2199   &   $ -$13.57   &   $ $126.5    &   $-$16.07   &  $-$14.45   &  17.6
\enddata
\tablecomments{\normalsize
Column (1) official SDSS name;
Column (2) luminosity of the power-law continuum at 3000\,\AA, $\lambda L_{\lambda}$(3000 \AA);
Column (3) FWHM of broad Mg\,II\,$\lambda2800$, corrected for instrumental broadening;
Column (4) flux of near-UV Fe\,II emission
(integrated in the range of 2200--3090 \AA\ from the best-fit model);
Column (5) rest-frame EW of UV Fe\,II emission;
Column (6) flux of the narrow component of Mg\,II\,$\lambda2800$;
Column (7) flux of the broad component of Mg\,II\,$\lambda2800$;
Column (8) rest-frame EW of the broad component of Mg\,II\,$\lambda2800$.
Luminosities, fluxes, EWs, and FWHM are in units of \lum, \flux, \AA, and \kms,
respectively. (This table is available in its entirety in a machine-readable form in the online
journal. A portion is shown here for guidance regarding its form and content.)}
\end{deluxetable}

\clearpage

\begin{table*}[h]
\topmargin 0.0cm
\evensidemargin = 0mm
\oddsidemargin = 0mm
\scriptsize 
\caption{Results of Spearman Correlation Analysis \tablenotemark{a} }
\label{corrtab}
\medskip
\vfill
\begin{tabular}{l|c c c c}
\hline \hline
   &
FWHM (\hb$^{\mathrm B}$)        &
$L_{5100}$\tablenotemark{b} &
\mbh                     \tablenotemark{b}  &
\lratio                  \tablenotemark{b}  \\
\hline
EW(Fe\,II$^{\mathrm N}\,\lambda4570$)                           &  $-$0.659 ( $<$1e-15 ) &   $~$0.102  ( $<$1e-15 ) &  $-$0.401 ( $<$1e-15 ) & $~$0.671 ( $<$1e-15 )   \\   
(Fe\,II$^{\rm N}\,\lambda4570$)/Mg\,II                          &  $-$0.731 ( $<$1e-15 ) &   $~$0.097  ( $~$8e-06 ) &  $-$0.521 ( $<$1e-15 ) & $~$0.695 ( $<$1e-15 )   \\   
(Fe\,II$^{\rm N}\,\lambda4570$)/\hb$^{\mathrm B}$               &  $-$0.680 ( $<$1e-15 ) &   $~$0.079  ( $~$2e-06 ) &  $-$0.417 ( $<$1e-15 ) & $~$0.668 ( $<$1e-15 )   \\   
(Fe\,II$^{\rm N}\,\lambda4570$)/[O\,III]\,$\lambda5007$         &  $-$0.606 ( $<$1e-15 ) &   $~$0.140  ( $<$1e-15 ) &  $-$0.319 ( $<$1e-15 ) & $~$0.626 ( $<$1e-15 )   \\   
EW(Fe\,II$^{\rm B}\,\lambda4570$)                               &  $-$0.333 ( $<$1e-15 ) &   $~$0.156  ( $<$1e-15 ) &  $-$0.150 ( $<$1e-15 ) & $~$0.398 ( $<$1e-15 )   \\   
(Fe\,II$^{\rm B}\,\lambda4570$)/Mg\,II                          &  $-$0.567 ( $<$1e-15 ) &   $~$0.116  ( $<$1e-15 ) &  $-$0.397 ( $<$1e-15 ) & $~$0.557 ( $<$1e-15 )   \\   
(Fe\,II$^{\rm B}\,\lambda4570$)/\hb$^{\mathrm B}$               &  $-$0.474 ( $<$1e-15 ) &   $~$0.064  ( $~$2e-05 ) &  $-$0.288 ( $<$1e-15 ) & $~$0.451 ( $<$1e-15 )   \\   
(Fe\,II$^{\rm B}\,\lambda4570$)/[O\,III]\,$\lambda5007$         &  $-$0.244 ( $<$1e-15 ) &   $~$0.174  ( $<$1e-15 ) &  $-$0.069 ( $~$8e-06 ) & $~$0.336 ( $<$1e-15 )   \\   
EW(Fe\,II\,UV)                                                  &  $~$0.020 ( $~$4e-01 ) &   $-$0.036  ( $~$7e-02 ) &  $-$0.018 ( $~$4e-01 ) & $-$0.044 ( $~$4e-02 )   \\   
(Fe\,II\,UV)/Mg\,II                                             &  $-$0.434 ( $<$1e-15 ) &   $~$0.157  ( $<$1e-15 ) &  $-$0.266 ( $<$1e-15 ) & $~$0.460 ( $<$1e-15 )   \\   
(Fe\,II\,UV)/\hb$^{\mathrm B}$                                  &  $-$0.161 ( $<$1e-15 ) &   $~$0.042  ( $~$5e-02 ) &  $-$0.112 ( $<$1e-15 ) & $~$0.158 ( $<$1e-15 )   \\   
(Fe\,II\,UV)/[O\,III]\,$\lambda5007$                            &  $-$0.068 ( $~$2e-03 ) &   $~$0.133  ( $<$1e-15 ) &  $-$0.024 ( $~$3e-01 ) & $~$0.136 ( $<$1e-15 )   \\   
(Fe\,II$^{\rm N}\,\lambda4570$)/(Fe\,II\,UV)                    &  $-$0.698 ( $<$1e-15 ) &   $~$0.060  ( $~$6e-03 ) &  $-$0.508 ( $<$1e-15 ) & $~$0.647 ( $<$1e-15 )   \\   
(Fe\,II$^{\rm B}\,\lambda4570$)/(Fe\,II\,UV)                    &  $-$0.434 ( $<$1e-15 ) &   $~$0.042  ( $~$5e-02 ) &  $-$0.322 ( $<$1e-15 ) & $~$0.401 ( $<$1e-15 )   \\   
(Fe\,II$^{\rm N}\,\lambda4570$)/(Fe\,II$^{\rm B}\,\lambda4570$) &  $-$0.580 ( $<$1e-15 ) &   $~$0.065  ( $~$2e-05 ) &  $-$0.364 ( $<$1e-15 ) & $~$0.585 ( $<$1e-15 )   \\   
EW([O\,III]\,$\lambda5007$)                                     &  $~$0.085 ( $~$4e-08 ) &   $-$0.128  ( $<$1e-15 ) &  $-$0.029 ( $~$6e-02 ) & $-$0.175 ( $<$1e-15 )   \\   
EW(\hb$^{\mathrm B}$)                                           &  $~$0.333 ( $<$1e-15 ) &   $~$0.110  ( $~$1e-12 ) &  $~$0.278 ( $<$1e-15 ) & $-$0.205 ( $<$1e-15 )   \\   
EW(Mg\,II)                                                      &  $~$0.496 ( $<$1e-15 ) &   $-$0.216  ( $<$1e-15 ) &  $~$0.280 ( $<$1e-15 ) & $-$0.552 ( $<$1e-15 )   \\   
\hline
\end{tabular}
\medskip
\vfill
{\normalsize $^a$~For each entry, we list the Spearman rank correlation
      coefficient (\rs) and the probability of the null hypothesis that
      the correlation is not present (\pnull) in parenthesis.
      For the correlations concerning UV emission lines, the data for the 2092 objects
      in the UV subsample are used; otherwise, those for the 4178 objects in the full sample are used.}\\
{\normalsize $^b$~$L_{5100} \equiv \lambda L_{\lambda}$(5100\,\AA);
the BH masses are calculated using
      the formalism presented in Wang et al. (2009, their Eqn.~11);
      Eddington ratios (\lratio) are calculated assuming
      that the bolometric luminosity $\lbol \approx 9\,L_{5100}$.}
\end{table*}
\clearpage

\begin{table*}[h]
\topmargin 0.0cm
\evensidemargin = 0mm
\oddsidemargin = 0mm
\tiny 
\caption{Results of Spearman Partial Correlation Analysis}
\label{tab-partialcorr}
\medskip
\vfill
\begin{tabular}{l|c c c c c c }
 \hline \hline
\backslashbox[25mm]{~~~~~~~X}{} &
(X\,,\,\lratio\,;\,FWHM(\hb$^{\mathrm B}$)\,)  &
(X\,,\,FWHM(\hb$^{\mathrm B}$);\,\lratio)      &
(X\,,\,\lratio\,;\,$L_{5100}$)                 &
(X\,,\,$L_{5100}$;\,\lratio)                   &
(X\,,\,\lratio\,;\,\mbh)                       &
(X\,,\,\mbh\,;\,\lratio)                       \\
\hline
EW(Fe\,II$^{\mathrm N}\,\lambda4570$)                           &  $~$0.391   (  $<$1e-15  )     &    $-$0.358  (  $<$1e-15  )  &  $ $0.727 ( $<$1e-15 )   &  $-$0.388 ( $<$1e-15  )          &    $~$0.661    (  $<$1e-15    )   &      $-$0.376    (   $<$1e-15    )    \\
(Fe\,II$^{\rm N}\,\lambda4570$)/Mg\,II                          &  $~$0.243   (  $<$1e-15  )     &    $-$0.394  (  $<$1e-15  )  &  $ $0.742 ( $<$1e-15 )   &  $-$0.374 ( $<$1e-15  )          &    $~$0.622    (  $<$1e-15    )   &      $-$0.373    (   $<$1e-15    )    \\
(Fe\,II$^{\rm N}\,\lambda4570$)/\hb$^{\mathrm B}$               &  $~$0.367   (  $<$1e-15  )     &    $-$0.400  (  $<$1e-15  )  &  $ $0.736 ( $<$1e-15 )   &  $-$0.421 ( $<$1e-15  )          &    $~$0.660    (  $<$1e-15    )   &      $-$0.397    (   $<$1e-15    )    \\
(Fe\,II$^{\rm N}\,\lambda4570$)/[O\,III]\,$\lambda5007$         &  $~$0.356   (  $<$1e-15  )     &    $-$0.302  (  $<$1e-15  )  &  $ $0.654 ( $<$1e-15 )   &  $-$0.277 ( $<$1e-15  )          &    $~$0.608    (  $<$1e-15    )   &      $-$0.261    (   $<$1e-15    )    \\
EW(Fe\,II$^{\rm B}\,\lambda4570$)                               &  $~$0.245   (  $<$1e-15  )     &    $-$0.084  (  $~$6e-08  )  &  $ $0.375 ( $<$1e-15 )   &  $-$0.064 ( $~$3e-05  )          &    $~$0.381    (  $<$1e-15    )   &      $-$0.082    (   $~$9e-08    )    \\
(Fe\,II$^{\rm B}\,\lambda4570$)/Mg\,II                          &  $~$0.196   (  $<$1e-15  )     &    $-$0.232  (  $<$1e-15  )  &  $ $0.575 ( $<$1e-15 )   &  $-$0.208 ( $<$1e-15  )          &    $~$0.475    (  $<$1e-15    )   &      $-$0.232    (   $<$1e-15    )    \\
(Fe\,II$^{\rm B}\,\lambda4570$)/\hb$^{\mathrm B}$               &  $~$0.190   (  $<$1e-15  )     &    $-$0.249  (  $<$1e-15  )  &  $ $0.490 ( $<$1e-15 )   &  $-$0.223 ( $<$1e-15  )          &    $~$0.421    (  $<$1e-15    )   &      $-$0.231    (   $<$1e-15    )    \\
(Fe\,II$^{\rm B}\,\lambda4570$)/[O\,III]\,$\lambda5007$         &  $~$0.238   (  $<$1e-15  )     &    $-$0.014  (  $~$4e-01  )  &  $ $0.292 ( $<$1e-15 )   &  $-$0.000 ( 1         )          &    $~$0.330    (  $<$1e-15    )   &      $-$0.006    (   $~$7e-01    )    \\
EW(Fe\,II\,UV)                                                  &  $-$0.048   (  $~$3e-02  )     &    $-$0.028  (  $~$2e-01  )  &  $-$0.030 ( $~$2e-01 )   &  $-$0.017 ( $~$4e-01  )          &    $-$0.056    (  $~$1e-02    )   &      $-$0.038    (   $~$8e-02    )    \\
(Fe\,II\,UV)/Mg\,II                                             &  $~$0.202   (  $<$1e-15  )     &    $-$0.112  (  $~$2e-07  )  &  $ $0.444 ( $<$1e-15 )   &  $-$0.082 ( $~$2e-04  )          &    $~$0.401    (  $<$1e-15    )   &      $-$0.103    (   $~$2e-06    )    \\
(Fe\,II\,UV)/\hb$^{\mathrm B}$                                  &  $~$0.046   (  $~$3e-02  )     &    $-$0.056  (  $~$1e-02  )  &  $ $0.157 ( $~$4e-13 )   &  $-$0.039 ( $~$7e-02  )          &    $~$0.125    (  $~$1e-08    )   &      $-$0.055    (   $~$1e-02    )    \\
(Fe\,II\,UV)/[O\,III]\,$\lambda5007$                            &  $~$0.140   (  $~$1e-10  )     &    $~$0.076  (  $~$5e-04  )  &  $ $0.083 ( $~$1e-04 )   &  $ $0.078 ( $~$4e-04  )          &    $~$0.138    (  $~$2e-10    )   &      $~$0.033    (   $~$1e-01    )    \\
(Fe\,II$^{\rm N}\,\lambda4570$)/(Fe\,II\,UV)                    &  $~$0.183   (  $<$1e-15  )     &    $-$0.384  (  $<$1e-15  )  &  $ $0.706 ( $<$1e-15 )   &  $-$0.374 ( $<$1e-15  )          &    $~$0.564    (  $<$1e-15    )   &      $-$0.360    (   $<$1e-15    )    \\
(Fe\,II$^{\rm B}\,\lambda4570$)/(Fe\,II\,UV)                    &  $~$0.088   (  $~$5e-05  )     &    $-$0.201  (  $<$1e-15  )  &  $ $0.434 ( $<$1e-15 )   &  $-$0.187 ( $<$1e-15  )          &    $~$0.315    (  $<$1e-15    )   &      $-$0.194    (   $<$1e-15    )    \\
(Fe\,II$^{\rm N}\,\lambda4570$)/(Fe\,II$^{\rm B}\,\lambda4570$) &  $~$0.308   (  $<$1e-15  )     &    $-$0.295  (  $<$1e-15  )  &  $ $0.646 ( $<$1e-15 )   &  $-$0.344 ( $<$1e-15  )          &    $~$0.564    (  $<$1e-15    )   &      $-$0.318    (   $<$1e-15    )    \\
EW([O\,III]\,$\lambda5007$)                                     &  $-$0.162   (  $<$1e-15  )     &    $-$0.053  (  $~$6e-04  )  &  $-$0.128 ( $<$1e-15 )   &  $-$0.044 ( $~$4e-03  )          &    $-$0.184    (  $<$1e-15    )   &      $-$0.064    (   $~$3e-05    )    \\
EW(\hb$^{\mathrm B}$)                                           &  $~$0.041   (  $~$8e-03  )     &    $~$0.271  (  $<$1e-15  )  &  $-$0.308 ( $<$1e-15 )   &  $ $0.259 ( $<$1e-15  )          &    $-$0.161    (  $<$1e-15    )   &      $~$0.249    (   $<$1e-15    )    \\
EW(Mg\,II)                                                      &  $-$0.292   (  $<$1e-15  )     &    $~$0.092  (  $~$3e-05  )  &  $-$0.523 ( $<$1e-15 )   &  $ $0.067 ( $~$2e-03  )          &    $-$0.500    (  $<$1e-15    )   &      $~$0.081    (   $~$2e-04    )    \\
\hline
\end{tabular}

\medskip
\vfill
{\normalsize Note. ---
      ($X$,\,$Y$;\,$Z$) denotes the partial correlation
      between $X$ and $Y$, controlling for $Z$.
      For each entry, we list the Spearman rank partial correlation
      coefficient (\rs) and the probability of the null hypothesis (\pnull) in parenthesis.
      For the correlations concerning UV emission lines, the data for the 2092 objects
      in the UV subsample are used; otherwise, those for the 4178 objects in the full sample are used.
      The AGN luminosities ($L_{5100} \equiv \lambda L_{\lambda}$(5100\,\AA)\,),
      BH masses and Eddington ratios (\lratio) are calculated in the same way as in Table \ref{corrtab}.}
\end{table*}
\clearpage


\begin{thebibliography}{}

\bibitem[Adelman-McCarthy et al.(2006)]{2006ApJS..162...38A}
Adelman-McCarthy, J.~K., et al.\ 2006, \apjs, 162, 38

\bibitem[Bachev et al.(2004)]{2004ApJ...617..171B} Bachev, R., Marziani,
P., Sulentic, J.~W., Zamanov, R., Calvani, M.,
\& Dultzin-Hacyan, D.\ 2004, \apj, 617, 171

\bibitem[Baldwin et al.(1995)]{1995ApJ...455L.119B} Baldwin, J., Ferland,
G., Korista, K., \& Verner, D.\ 1995, \apjl, 455, L119  

\bibitem[Baldwin(1977)]{1977ApJ...214..679B} Baldwin, J.~A.\ 1977, \apj,
214, 679

\bibitem[Baldwin et al.(2004)]{2004ApJ...615..610B} Baldwin, J.~A.,
Ferland, G.~J., Korista, K.~T., Hamann, F., \& LaCluyz{\'e}, A.\ 2004, \apj, 615, 610

\bibitem[Balogh et al.(1999)]{1999ApJ...527...54B} Balogh, M.~L., Morris,
S.~L., Yee, H.~K.~C., Carlberg, R.~G.,
\& Ellingson, E.\ 1999, \apj, 527, 54


\bibitem[Barth et al.(2003)]{2003ApJ...594L..95B} Barth, A.~J., Martini,
P., Nelson, C.~H., \& Ho, L.~C.\ 2003, \apjl, 594, L95


\bibitem[Barvainis(1987)]{1987ApJ...320..537B} Barvainis, R.\ 1987, \apj, 320, 537


\bibitem[Baskin \& Laor(2004)]{2004MNRAS.350L..31B} Baskin, A., \& Laor, A.\ 2004, \mnras, 350, L31


\bibitem[Becker et al.(2000)]{2000ApJ...538...72B} Becker, R.~H., White,
R.~L., Gregg, M.~D., Brotherton, M.~S., Laurent-Muehleisen, S.~A.,
\& Arav, N.\ 2000, \apj, 538, 72

\bibitem[Bentz et al.(2009)]{2009ApJ...697..160B} Bentz, M.~C., Peterson, B.~M., Netzer, H., Pogge, R.~W., \& Vestergaard, M.\ 2009, \apj, 697, 160

\bibitem[Boroson(2002)]{2002ApJ...565...78B} Boroson, T.~A.\ 2002, \apj, 565, 78

\bibitem[Boroson \& Green(1992)]{1992ApJS...80..109B} Boroson, T.~A.~\&
Green, R.~F.\ 1992, \apjs, 80, 109

\bibitem[Boroson et al.(1985)]{1985ApJ...293..120B} Boroson, T.~A.,
Persson, S.~E., \& Oke, J.~B.\ 1985, \apj, 293, 120


\bibitem[Brotherton et al.(1994)]{1994ApJ...430..495B} Brotherton, M.~S.,
Wills, B.~J., Francis, P.~J., \& Steidel, C.~C.\ 1994, \apj, 430, 495

\bibitem[Bruhweiler \& Verner(2008)]{2008ApJ...675...83B} Bruhweiler, F., \& Verner, E.\ 2008, \apj, 675, 83

\bibitem[Chen et al.(1989)]{1989ApJ...339..742C} Chen, K., Halpern, J.~P.,
\& Filippenko, A.~V.\ 1989, \apj, 339, 742

\bibitem[Collin et al.(2006)]{2006A&A...456...75C} Collin, S., Kawaguchi, T.,
Peterson, B.~M., \& Vestergaard, M.\ 2006, \aap, 456, 75

\bibitem[Collin \& Hur{\'e}(2001)]{2001A&A...372...50C} Collin, S., \& Hur{\'e}, J.-M.\ 2001, \aap, 372, 50

\bibitem[Collin \& Joly(2000)]{2000NewAR..44..531C} Collin, S., \& Joly, M.\ 2000, NewAR, 44, 531

\bibitem[Collin-Souffrin(1987)]{1987A&A...179...60C} Collin-Souffrin, S.\ 1987, \aap, 179, 60

\bibitem[Collin-Souffrin et al.(1980)]{1980A&A....83..190C} Collin-Souffrin, S., Joly, M., Dumont, S., \& Heidmann, N.\ 1980, \aap, 83, 190


\bibitem[Collin-Souffrin et
al.(1986)]{1986A&A...166...27C} Collin-Souffrin, S., Joly, M., P\'equignot, D., \& Dumont, S.\ 1986, \aap, 166, 27

\bibitem[Davies et al.(2007)]{2007ApJ...671.1388D} Davies, R.~I., M\"uller
S{\'a}nchez, F., Genzel, R., Tacconi, L.~J., Hicks, E.~K.~S., Friedrich,
S., \& Sternberg, A.\ 2007, \apj, 671, 1388

\bibitem[Dietrich et al.(2003)]{2003ApJ...596..817D} Dietrich, M., Hamann,
F., Appenzeller, I., \& Vestergaard, M.\ 2003, \apj, 596, 817

\bibitem[Dong et al.(2010)]{Dong10}
Dong,~X.-B., Ho,~L.~C., Wang,~J.-G., Wang,~T.-G., Wang,~H., Fan,~X., \& Zhou,~H.
2010, \apjl, 721, L143

\bibitem[Dong et al.(2009b)]{2009ASPC..408...83D} Dong, X.-B., Wang, J.-G.,
Wang, T.-G., Wang, H., Fan, X., Zhou, H., Yuan, W., \& Long, Q.\ 2009b,
in ASP Conf. Ser. 408, The Starburst-AGN Connection, ed. W. Wang et al.
(San Francisco, CA: ASP), 83

\bibitem[Dong et al.(2009a)]{2009ApJ...703L...1D} Dong, X.-B., Wang, T.-G.,
Wang, J.-G., Fan, X., Wang, H., Zhou, H., \& Yuan, W.\ 2009a, \apjl, 703, L1

\bibitem[Dong et al.(2008)]{2008MNRAS.383..581D} Dong, X., Wang, T., Wang, J., Yuan, W., Zhou, H., Dai, H., \& Zhang, K.\ 2008, \mnras, 383, 581

\bibitem[Dong et al.(2005)]{2005ApJ...620..629D} Dong, X.-B., Zhou, H.-Y.,
Wang, T.-G., Wang, J.-X., Li, C., \& Zhou, Y.-Y.\ 2005, \apj, 620, 629

\bibitem[Elitzur \& Shlosman(2006)]{2006ApJ...648L.101E}
Elitzur, M., \& Shlosman, I.\ 2006, \apjl, 648, L101

\bibitem[Elston et al.(1994)]{1994Natur.367..250E} Elston, R., Thompson,
K.~L., \& Hill, G.~J.\ 1994, \nat, 367, 250

\bibitem[Elvis et al.(1994)]{1994ApJS...95....1E} Elvis, M., et al.\ 1994,
\apjs, 95, 1

\bibitem[Eracleous \& Halpern(2003)]{2003ApJ...599..886E} Eracleous, M., \&
Halpern, J.~P.\ 2003, \apj, 599, 886

\bibitem[Fabian et al.(2006)]{2006MNRAS.373L..16F} Fabian, A.~C., Celotti,
A., \& Erlund, M.~C.\ 2006, \mnras, 373, L16

\bibitem[Fabian et al.(2009)]{2009MNRAS.394L..89F} Fabian, A.~C.,
Vasudevan, R.~V., Mushotzky, R.~F., Winter, L.~M.,
\& Reynolds, C.~S.\ 2009, \mnras, 394, L89

\bibitem[Ferland et al.(2009)]{2009ApJ...707L..82F} Ferland, G.~J., Hu, C.,
Wang, J.-M., Baldwin, J.~A., Porter, R.~L., van Hoof, P.~A.~M.,
\& Williams, R.~J.~R.\ 2009, \apjl, 707, L82

\bibitem[Ferland et al.(1998)]{1998PASP..110..761F} Ferland, G.~J.,
Korista, K.~T., Verner, D.~A., Ferguson, J.~W., Kingdon, J.~B.,
\& Verner, E.~M.\ 1998, \pasp, 110, 761

\bibitem[Ferland et al.(1992)]{Ferland92} Ferland, G.~J., Peterson, B.~M., Horne, K., Welsh, W.~F., \& Nahar, S.~N.  1992, \apj, 387, 95


\bibitem[Fitzpatrick(1999)]{1999PASP..111...63F} Fitzpatrick, E.~L.\ 1999,
\pasp, 111, 63

\bibitem[Freudling et al.(2003)]{2003ApJ...587L..67F} Freudling, W.,
Corbin, M.~R., \& Korista, K.~T.\ 2003, \apjl, 587, L67

\bibitem[Gaskell \& Goosmann(2008)]{blr-torus} Gaskell, C.~M., \& Goosmann, R.~W.\ 2008, arXiv:0805.4258

\bibitem[Granato \& Danese(1994)]{1994MNRAS.268..235G} Granato, G.~L., \& Danese, L.\ 1994, \mnras, 268, 235

\bibitem[Grupe et al.(2010)]{2010ApJS..187...64G} Grupe, D., Komossa, S.,
Leighly, K.~M., \& Page, K.~L.\ 2010, \apjs, 187, 64

\bibitem[Halpern et al.(1996)]{1996ApJ...464..704H} Halpern, J.~P.,
Eracleous, M., Filippenko, A.~V., \& Chen, K.\ 1996, \apj, 464, 704

\bibitem[Hamann \& Ferland(1993)]{1993ApJ...418...11H} Hamann, F., \& Ferland, G.\ 1993, \apj, 418, 11

\bibitem[Ho(2008)]{Ho08} Ho, L.~C. 2008, ARA\&A, 46, 475

\bibitem[Hu et al.(2008a)]{Hu08a} Hu, C., Wang, J.-M., Ho, L. C., Chen, Y.-M., Bian, W.-H., \& Xue, S.-J. 2008a, \apj, 683, L115

\bibitem[Hu et al.(2008b)]{2008ApJ...687...78H} Hu, C., Wang, J.-M., Ho,
L.~C., Chen, Y.-M., Zhang, H.-T., Bian, W.-H.,
\& Xue, S.-J.\ 2008b, \apj, 687, 78

\bibitem[Isobe et al.(1986)]{1986ApJ...306..490I} Isobe, T., Feigelson,
E.~D., \& Nelson, P.~I.\ 1986, \apj, 306, 490

\bibitem[Iwamuro et al.(2004)]{2004ApJ...614...69I} Iwamuro, F., Kimura,
M., Eto, S., Maihara, T., Motohara, K., Yoshii, Y.,
\& Doi, M.\ 2004, \apj, 614, 69

\bibitem[Jaffe et al.(2004)]{2004Natur.429...47J} Jaffe, W., et al.\ 2004, \nat, 429, 47

\bibitem[Jiang et al.(2007)]{2007AJ....134.1150J} Jiang, L., Fan, X., Vestergaard, M., Kurk, J.~D., Walter, F., Kelly, B.~C., \& Strauss, M.~A.\ 2007, \aj, 134, 1150


\bibitem[Joly et al.(2008)]{2008RMxAC..32...59J}
Joly, M., V{\'e}ron-Cetty, M., \& V{\'e}ron, P.\ 2008, Rev. Mexicana Astron. Astrofis. Conf. Ser., 32, 59

\bibitem[Kaspi et al.(2000)]{2000ApJ...533..631K} Kaspi, S., Smith, P.~S.,
Netzer, H., Maoz, D., Jannuzi, B.~T., \& Giveon, U.,\ 2000, \apj,
533, 631

\bibitem[Kelly(2007)]{2007ApJ...665.1489K} Kelly, B.~C.\ 2007, \apj, 665,
1489

\bibitem[Kendall \& Stuart(1979)]{1979ats..book.....K}
Kendall, M., \& Stuart, A.\ 1979,
The Advanced Theory of Statistics. Vol.2: Inference and Relationship, 4th Edition (London: Griffin)

\bibitem[Kl{\"o}ckner et al.(2003)]{2003Natur.421..821K} Kl{\"o}ckner,
H.-R., Baan, W.~A., \& Garrett, M.~A.\ 2003, \nat, 421, 821


\bibitem[Kollatschny \& Welsh(2001)]{2001ASPC..224..449K}
Kollatschny, W., \& Welsh, W.~F.\ 2001, in ASP Conf. Ser. 224,
Probing the Physics of Active Galactic Nuclei, ed. B.~M. Peterson,
R.~S. Polidan, \& R.~W. Pogge (San Francisco, CA: ASP), 449


\bibitem[Korista et al.(1998)]{1998ApJ...507...24K} Korista, K., Baldwin,
J., \& Ferland, G.\ 1998, \apj, 507, 24

\bibitem[Korista(1999)]{1999ASPC..162..429K} Korista, K.\ 1999,
in ASP Conf. Ser. 162, Quasars and Cosmology,
ed. G.~Ferland \& J. Baldwin (San Francisco, CA: ASP), 429


\bibitem[Kova{\v c}evi{\'c} et al.(2010)]{2010ApJS..189...15K} Kova{\v
c}evi{\'c}, J., Popovi{\'c}, L.~{\v C}.,
\& Dimitrijevi{\'c}, M.~S.\ 2010, \apjs, 189, 15

\bibitem[Kuehn et al.(2008)]{2008ApJ...673...69K} Kuehn, C.~A., Baldwin,
J.~A., Peterson, B.~M., \& Korista, K.~T.\ 2008, \apj, 673, 69

\bibitem[Kurk et al.(2007)]{2007ApJ...669...32K} Kurk, J.~D., et al.\ 2007,
\apj, 669, 32

\bibitem[Landt et al.(2008)]{Landt08} Landt, H.,, Bentz, M. C., Ward, M. J., Elvis, M., Peterson, B. M., Korista, K. T., \& Karovska, M. 2008, \apjs, 174, 282

\bibitem[Laor(2007)]{2007ASPC..373..384L} Laor, A.\ 2007, in The Central Engine of Active Galactic Nuclei, ed. L. C. Ho \& J.-M. Wang (San Francisco: ASP), 384

\bibitem[Laor et al.(1994)]{1994ApJ...420..110L} Laor, A., Bahcall, J.~N.,
Jannuzi, B.~T., Schneider, D.~P., Green, R.~F.,
\& Hartig, G.~F.\ 1994, \apj, 420, 110

\bibitem[Laor et al.(1997)]{1997ApJ...489..656L} Laor, A., Jannuzi, B.~T.,
Green, R.~F., \& Boroson, T.~A.\ 1997, \apj, 489, 656

\bibitem[Leighly(2004)]{2004ApJ...611..125L} Leighly, K.~M.\ 2004, \apj,
611, 125

\bibitem[Leighly \& Moore(2006)]{2006ApJ...644..748L} Leighly, K.~M., \& Moore, J.~R.\ 2006, \apj, 644, 748

\bibitem[Macklin(1982)]{Macklin82} Macklin, J.~T. 1982, \mnras, 199, 1119

\bibitem[Marconi et al.(2008)]{2008ApJ...678..693M} Marconi, A., Axon,
D.~J., Maiolino, R., Nagao, T., Pastorini, G., Pietrini, P., Robinson, A.,
\& Torricelli, G.\ 2008, \apj, 678, 693

\bibitem[Markwardt(2009)]{2009ASPC..411..251M} Markwardt, C.~B.\ 2009, in
Astronomical Data Analysis Software and Systems XVIII,
ed. D. A. Bohlender, D. Durand, \& P. Dowler (San Francisco: ASP), 251

\bibitem[Marziani et al.(2003)]{2003ApJS..145..199M} Marziani, P.,
Sulentic, J.~W., Zamanov, R., Calvani, M., Dultzin-Hacyan, D., Bachev, R.,
\& Zwitter, T.\ 2003, \apjs, 145, 199


\bibitem[Matsuoka et al.(2007)]{2007ApJ...663..781M} Matsuoka, Y., Oyabu,
S., Tsuzuki, Y., \& Kawara, K.\ 2007, \apj, 663, 781

\bibitem[Matteucci
\& Recchi(2001)]{2001ApJ...558..351M} Matteucci, F., \& Recchi, S.\ 2001, \apj, 558, 351

\bibitem[McLure \& Dunlop(2004)]{MD04} McLure, R.~J., \& Dunlop, J.~S. 2004, \mnras, 352, 1390

\bibitem[Merloni \& Heinz(2008)]{2008MNRAS.388.1011M} Merloni, A., \& Heinz, S.\ 2008, \mnras, 388, 1011

\bibitem[Murray \& Chiang(1997)]{1997ApJ...474...91M} Murray, N., \& Chiang, J.\ 1997, \apj, 474, 91

\bibitem[Murray et al.(1995)]{1995ApJ...451..498M} Murray, N., Chiang, J.,
Grossman, S.~A., \& Voit, G.~M.\ 1995, \apj, 451, 498

\bibitem[Netzer(1985)]{1985MNRAS.216...63N} Netzer, H.\ 1985, \mnras, 216, 63



\bibitem[Nagao et al.(2003)]{2003AJ....126.1167N} Nagao, T., Murayama, T.,
Shioya, Y., \& Taniguchi, Y.\ 2003, \aj, 126, 1167

\bibitem[Osterbrock \& Ferland(2006)]{2006agna.book.....O}
Osterbrock, D.~E., \& Ferland, G.~J.\ 2006,
Astrophysics of Gaseous Nebulae and Active Galactic Nuclei, 2nd~ed.~
(Sausalito, CA: University Science Books)


\bibitem[Proga et al.(2008)]{2008ApJ...676..101P} Proga, D., Ostriker,
J.~P., \& Kurosawa, R.\ 2008, \apj, 676, 101


\bibitem[Roelofs et al.(2006)]{2006MNRAS.365.1109R} Roelofs, G.~H.~A.,
Groot, P.~J., Marsh, T.~R., Steeghs, D.,
\& Nelemans, G.\ 2006, \mnras, 365, 1109


\bibitem[Salviander et al.(2007)]{2007ApJ...662..131} Salviander, S., Shields, G. A., Gebhardt, K., \& Bonning, E. W. 2007, \apj, 662, 131

\bibitem[Sameshima et al.(2011)]{2011MNRAS.410.1018S} Sameshima, H.,
Kawara, K., Matsuoka, Y., Oyabu, S., Asami, N.,
\& Ienaka, N.\ 2011, \mnras, 410, 1018


\bibitem[Sanders et al.(1988)]{1988ApJ...325...74S} Sanders, D.~B., Soifer,
B.~T., Elias, J.~H., Madore, B.~F., Matthews, K., Neugebauer, G.,
\& Scoville, N.~Z.\ 1988, \apj, 325, 74

\bibitem[Schlegel, Finkbeiner, \& Davis(1998)]{sfd98}
Schlegel, D.~J., Finkbeiner, D.~P., \& Davis, M.\ 1998, \apj, 500, 525


\bibitem[Schmitt et al.(2003)]{2003ApJ...597..768S} Schmitt, H.~R., Donley,
J.~L., Antonucci, R.~R.~J., Hutchings, J.~B., Kinney, A.~L.,
\& Pringle, J.~E.\ 2003, \apj, 597, 768



\bibitem[Shakura \& Syunyaev(1973)]{1973A&A....24..337S} Shakura, N.~I., \& Sunyaev, R.~A.\ 1973, \aap, 24, 337

\bibitem[Shemmer et al.(2004)]{2004ApJ...614..547S} Shemmer, O., Netzer,
H., Maiolino, R., Oliva, E., Croom, S., Corbett, E.,
\& di Fabrizio, L.\ 2004, \apj, 614, 547

\bibitem[Shields et al.(1995)]{Shields95} Shields, J.~C., Ferland, G.~J., \& Peterson, B.~M. 1995, \apj, 441, 507


\bibitem[Smith et al.(2006)]{2006MNRAS.369.1537S} Smith, A.~J., Haswell,
C.~A., \& Hynes, R.~I.\ 2006, \mnras, 369, 1537

\bibitem[Strateva et al.(2003)]{2003AJ....126.1720S} Strateva, I.~V., et
al.\ 2003, \aj, 126, 1720


\bibitem[Suganuma et al.(2006)]{2006ApJ...639...46S} Suganuma, M., et al.\
2006, \apj, 639, 46


\bibitem[Sulentic et
al.(2000)]{2000ARA&A..38..521S} Sulentic, J.~W., Marziani, P., \& Dultzin-Hacyan, D.\ 2000a, \araa, 38, 521

\bibitem[Sulentic et al.(2000)]{2000ApJ...536L...5S} Sulentic, J.~W.,
Zwitter, T., Marziani, P., \& Dultzin-Hacyan, D.\ 2000b, \apjl, 536, L5

\bibitem[Tsuzuki et al.(2006)]{2006ApJ...650...57T} Tsuzuki, Y., Kawara,
K., Yoshii, Y., Oyabu, S., Tanab{\'e}, T.,
\& Matsuoka, Y.\ 2006, \apj, 650, 57


\bibitem[Vasudevan \& Fabian(2009)]{2009MNRAS.392.1124V} Vasudevan, R.~V., \& Fabian, A.~C.\ 2009, \mnras, 392, 1124


\bibitem[V{\'e}ron-Cetty et al.(2004)]{2004A&A...417..515V}
V{\'e}ron-Cetty, M.-P., Joly, M., \& V{\'e}ron, P.\ 2004, \aap, 417, 515

\bibitem[V{\'e}ron-Cetty et al.(2006)]
{2006A&A...451..851V} V{\'e}ron-Cetty, M.-P., Joly, M., V{\'e}ron, P., Boroson, T., Lipari, S., \& Ogle, P.\ 2006, \aap, 451, 851


\bibitem[V{\'e}ron-Cetty et al.(2007)]{2007A&A...475..487V} V{\'e}ron-Cetty, M.-P., V{\'e}ron, P., Joly, M., \& Kollatschny, W.\ 2007, \aap, 475, 487


\bibitem[Verner et al.(2004)]{2004ApJ...611..780V} Verner, E., Bruhweiler,
F., Verner, D., Johansson, S., Kallman, T.,
\& Gull, T.\ 2004, \apj, 611, 780

\bibitem[Verner et al.(2000)]{Verner00} Verner, E.~M., Verner, D.~A., Baldwin, J.~A., Ferland, G.~J., \& Martin, P.~G. 2000, \apj, 543, 831

\bibitem[Vestergaard \& Osmer(2009)]{VO} Vestergaard, M., \& Osmer, P. S. 2009, \apj, 699, 800

\bibitem[Vestergaard \& Peterson(2005)]{2005ApJ...625..688V} Vestergaard, M., \& Peterson, B.~M.\ 2005, \apj, 625, 688

\bibitem[Vestergaard \& Peterson(2006)]{2006ApJ...641..689V} Vestergaard,
M., \& Peterson, B.~M.\ 2006, \apj, 641, 689

\bibitem[Vestergaard \& Wilkes(2001)]{2001ApJS..134....1V} Vestergaard, M., \& Wilkes, B.~J.\ 2001, \apjs, 134, 1

\bibitem[Wandel(1999)]{1999ApJ...527..649W} Wandel, A.\ 1999, \apj, 527,
649




\bibitem[Wang et al.(2005)]{2005A&A...436..417W}
Wang, J., Wei, J.~Y., \& He, X.~T.\ 2005, \aap, 436, 417


\bibitem[Wang et al.(2009)]{2009ApJ...707.1334W}
Wang, J.-G., et al. 2009, \apj, 707, 1334

\bibitem[Wang et al.(2008)]{2008ApJ...674..668W} Wang, T., Dai, H.,
\& Zhou, H.\ 2008, \apj, 674, 668

\bibitem[Wang et al.(2005)]{2005ApJ...625L..35W} Wang, T.-G., Dong, X.-B.,
Zhang, X.-G., Zhou, H.-Y., Wang, J.-X., \& Lu, Y.-J.\ 2005, \apjl, 625, L35

\bibitem[Wang et
al.(1996)]{1996A&A...309...81W} Wang, T., Brinkmann, W., \& Bergeron, J.\ 1996, \aap, 309, 81



\bibitem[Warner et al.(2004)]{2004ApJ...608..136W} Warner, C., Hamann, F.,
\& Dietrich, M.\ 2004, \apj, 608, 136



\bibitem[Wu et al.(2008)]{2008MNRAS.389..213W} Wu, S.-M., Wang, T.-G.,
\& Dong, X.-B.\ 2008, \mnras, 389, 213

\bibitem[York et al.(2000)]{2000AJ....120.1579Y} York, D.~G., et al.\ 2000,
\aj, 120, 1579

\bibitem[Zhang et al.(2009)]{2009ASPC..408..281Z} Zhang, K., Wang, T.-G.,
Dong, X.-B., Zhou, H.-Y., \& Lu, H.-L.\ 2009,
in ASP Conf. Ser. 408, The Starburst-AGN Connection,
ed. W. Wang et al. (San Francisco, CA: ASP), 281

\bibitem[Zhang et al.(2008)]{2008ApJ...685L.109Z} Zhang, K., Wang, T.,
Dong, X., \& Lu, H.\ 2008, \apjl, 685, L109

\bibitem[Zhang et al.(2010)]{2010ApJ...714..367Z} Zhang, S., Wang, T.-G.,
Wang, H., Zhou, H., Dong, X.-B., \& Wang, J.-G.\ 2010, \apj, 714, 367

\bibitem[Zhang et al.(2006)]{2006MNRAS.372L...5Z} Zhang, X.-G.,
Dultzin-Hacyan, D., \& Wang, T.-G.\ 2006, \mnras, 372, L5

\bibitem[Zhang et al.(2007)]{2007RMxAA..43..101Z} Zhang, X.-G.,
Dultzin-Hacyan, D., \& Wang, T.-G.\ 2007, Rev. Mexicana Astron. Astrofis., 43, 101


\bibitem[Zheng \& Malkan(1993)]{1993ApJ...415..517Z}
Zheng, W., \& Malkan, M.~A.\ 1993, \apj, 415, 517

\bibitem[Zhou et al.(2006)]{2006ApJS..166..128Z} Zhou, H., Wang, T., Yuan,
W., Lu, H., Dong, X., Wang, J., \& Lu, Y.\ 2006, \apjs, 166, 128


\end{thebibliography}
\end{document}